\documentclass[10pt, conference, letterpaper]{IEEEtran}
\IEEEoverridecommandlockouts
\usepackage{cite}
\usepackage{amsmath,amssymb,amsfonts}
\usepackage{algorithmic}
\usepackage{algorithm}
\usepackage{graphicx}
\usepackage{textcomp}
\usepackage{xcolor}
\usepackage{amsmath}  
\usepackage{adjustbox}
\usepackage{relsize}
\usepackage{caption}
\usepackage{subfig}
\usepackage{url}
\usepackage{dsfont}

\newcommand{\graphicsextension}{png}

\def\BibTeX{{\rm B\kern-.05em{\sc i\kern-.025em b}\kern-.08em
    T\kern-.1667em\lower.7ex\hbox{E}\kern-.125emX}}

\begin{document}

\title{Edge-Served Congestion Control for Wireless Multipath Transmission with a Transformer Agent}

\author{
\IEEEauthorblockN{Liang Wang}
\IEEEauthorblockA{
\textit{School of Electronic and Information Engineering} \\
\textit{Beijing Jiaotong University} \\
Beijing, China \\
Email: wangliang1@bjtu.edu.cn}
}

\maketitle

\begin{abstract}
  Multipath TCP is widely adopted to enhance connection quality-of-service by leveraging multiple network pathways on modern devices. However, the evolution of its core congestion control is hindered by the OS kernel, whose monolithic design imposes high development overhead and lacks the resource flexibility required for data-driven methods. Furthermore, inherent noise in network statistics induces a partial observability problem, which can mislead data-driven methods like Deep Reinforcement Learning. To bridge this gap, we propose Jazz, a system that re-architects multipath congestion control through a decoupled architecture that separates the decision-making ``brain'' from the in-kernel datapath, enabling it to operate on an external (edge) entity. At its core, Jazz employs a Transformer-based agent that processes sequences of historical observations to overcome the partial observability of single-step reinforcement learning. This allows it to learn and master fluctuating link conditions and intricate cross-path dependencies. Tested on a dual-band (5GHz/6GHz) Wi-Fi testbed, our implementation improves bandwidth efficiency by at least 2.85\% over conventional methods and maintains 96.2\% performance under 1\% packet loss, validating this design as a practical blueprint for agile network intelligence. 
\end{abstract}

\begin{IEEEkeywords}
Multipath TCP, Congestion Control, Edge Computing, Transformer-based Reinforcement Learning 
\end{IEEEkeywords}

\section{Introduction}
\label{section:1}
The proliferation of multi-homed devices, from mobile clients to datacenter servers, has driven the development of multipath transport protocols. Multipath TCP (MPTCP), standardized by the IETF in RFC 8684 \cite{rfc8684}, has emerged as the leading solution. Its integration into mainstream operating systems, including the Linux kernel and Apple's iOS, underscores its maturity and widespread deployment. The efficacy of MPTCP fundamentally hinges on its congestion control (CC) mechanism, which orchestrates data transmission across disparate paths to form a single, resilient, and high-throughput connection. It must dynamically adapt to the diverse and time-varying characteristics of each subflow to achieve ideal bandwidth utilization and queueing depth. The advent of data-driven modeling and large-scale analytics opens new avenues for more efficient multipath CC methods. However, these technologies demand a level of resource flexibility that traditional in-kernel datapaths were never designed to accommodate.




The traditional architecture of network functions poses a multifaceted challenge to network innovation \cite{narayan2018restructuring}. Firstly, the system kernel imposes great developmental barriers. Its stringent programming environment, including challenges with limited debugging tools and the complexity of managing concurrency, demands specialized expertise and lengthy development cycles. Secondly, the kernel presents a resource-constrained sandbox, ill-suited for the complex computational models that rely on features (e.g., floating-point arithmetic) unavailable in such an environment. This limitation also precludes access to standard user-space libraries and dedicated hardware accelerators like GPUs. Additionally, kernel-level intricate implementations carry inherent risks; minor errors can readily cause system-wide failures and critical vulnerabilities.

The tension between kernel stability and rapid functional evolution has spurred an industry-wide trend of moving network functions out of the monolithic kernel to enhance datapath programmability, enabling innovation on transport logic at the speed of software, not OS releases. The emergence of kernel-bypass technologies like DPDK \cite{intel_dpdk_2014} and user-space protocols like QUIC \cite{Iyengar2021QUIC} are testaments to this need, though they trade full compatibility with the native, standard TCP/IP stack for this agility. These approaches exemplify a broader principle: separating complex, evolving control logic from the static, performance-critical packet forwarding plane. This very principle of separation creates a fertile ground for advanced, data-driven methodologies.

Many studies have explored data-driven methods in CC. The efficacy of Deep Reinforcement Learning (DRL) stands out for its ability to autonomously learn a control policy that maps real-time network metrics to control decisions \cite{Xiao2021, Giacomoni2024}. The policy is progressively optimized through environmental interaction, allowing it to adapt to dynamic conditions where conventional approaches often fall short. While promising, our empirical findings indicate a key constraint in single-step RL like Deep Q Network (DQN) \cite{dqn}. Although capable of learning near-optimal policies from network conditions and feedback, their single-step inputs limit fine-grained control actions (further discussed in Section \ref{sec:motive}). To overcome this limitation, we employ an agent based on Transformer \cite{attention} that processes sequences of state observations, leveraging its ability to comprehend historical context to facilitate faster and more accurate real-time control.

To address these challenges, we propose Jazz, a system that enables edge-side online policy learning to adaptively manage CC for end-host multipath connections. First, to overcome the rigidity of kernel-centric network functions while ensuring full compatibility with the existing TCP/IP stack, we decouple the MPTCP CC logic from its in-kernel execution by implementing a minimal client that enforces directives from a decision engine operating outside the kernel. This separation enables control logic offloading to an edge-side entity. Second, to address the intricate link dynamics and inter-subflow dependencies inherent in multipath transmission, we engineer a Transformer-based CC agent. By processing sequences of historical observations, this agent overcomes the partial observability that limits single-step RL methods. It learns to discern underlying network trends from transient noisy metrics, enabling a holistic and adaptive control policy that coordinates all subflows. The contributions of this paper are summarized as follows:
\begin{itemize}
\item To the best of our knowledge, Jazz represents the first Transformer-based CC method deployable in real-world network infrastructure.
\item We design a framework that decouples decision-making from the constrained datapath, enabling agile development and deployment of network control strategies on an edge-side engine.
\item We develop a Transformer-based CC agent to tackle the challenge of partial observability in interpreting noisy network metrics. It processes sequences of observations to discern network dynamics and cross-link relationships.
\item We validate Jazz's effectiveness through experiments in both simulated topologies and a dual-band (5GHz/6GHz) Wi-Fi testbed. The results demonstrate robust performance across diverse conditions, and confirm the feasibility of edge-based control through systematic comparison between local and edge deployments. 
\end{itemize}

The remainder of the paper covers our motivation (Section \ref{sec:motive}), the conceptual framework, problem formulation, resulting system design (Sections \ref{sec:framework}-\ref{sec:system}), validation (Section \ref{sec:validation}), related work (Section \ref{sec:related}), and conclusion (Section \ref{sec:conclusion}).

\section{Motivation}
\label{sec:motive}
This section empirically illustrates the motivation for our design by examining three questions: first, the rationale for employing DRL methods; second, the partial observability problem that limits single-step RL; and third, how the Transformer overcomes this challenge for more robust control. 
\subsection{From Traditional Algorithm to Learning-based Method}
\begin{figure}[H]
  \vspace{-0.2cm}
  \centering
  \includegraphics[width=\columnwidth]{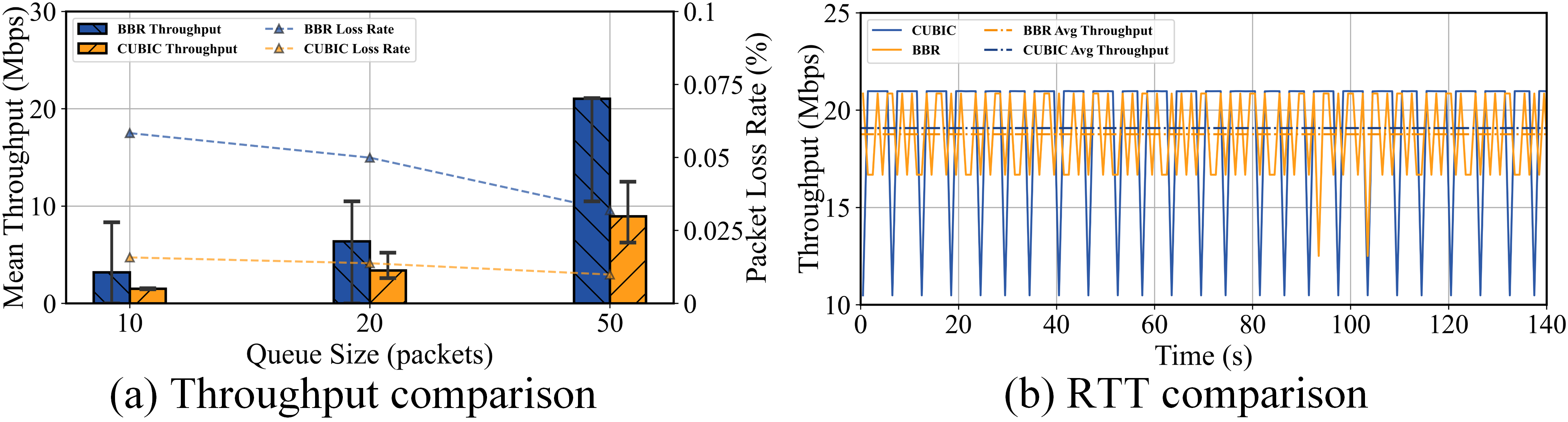}
  \caption
  {Performance comparison between CUBIC and BBR. The queue buffer was simulated using the Linux Traffic Control (TC). During the test in (b), the buffer size was set to 1000 packets (greater than the actual requirement).}
  \label{fig1}
  \vspace{-0.2cm}
  \end{figure}

Traditional CC algorithms are challenged by two key issues. First, they mostly treat each subflow as an independent TCP flow, which makes it difficult to manage the coupled interference where actions on one path negatively impact others. Second, their reliance on fixed heuristics limits their effectiveness when adapting to highly dynamic wireless links, making it challenging for any algorithm to perform consistently well across diverse network environments. For instance, when transferring 1GB of data between two hosts in our first scenario (Fig. \ref{fig1}(a)), we configured a 200 Mbps link with 60ms delay, 0.01\% packet loss, and a shallow buffer. Here, BBR \cite{cardwell2016bbr} delivered higher throughput but at the cost of greater packet loss. In a second scenario emulating a small Bandwidth-Delay Product (BDP) network (20 Mbps, 5ms delay, sufficient buffer), Fig. \ref{fig1}(b) shows that CUBIC \cite{ha2008cubic} achieved superior throughput yet exhibited significant fluctuations in both bandwidth and RTT. These results underscore that an algorithm's performance in terms of delivery rate, delay, and stability is highly contingent on the specific network environment.
DRL addresses these limitations by learning a global control policy for the entire connection directly from environmental interaction that effectively optimizes for inter-link dependencies. It can learn optimal strategies directly from experience rather than relying on predefined network models, enabling adaptation to various network conditions. Furthermore, the control objective in DRL is explicitly defined through a customizable reward function, allowing its behavior to be easily fine-tuned for varied demands, such as prioritizing low latency or maximizing throughput.
  
\begin{figure}[H]
\vspace{-0.1cm}
\centering
\includegraphics[width=\columnwidth]{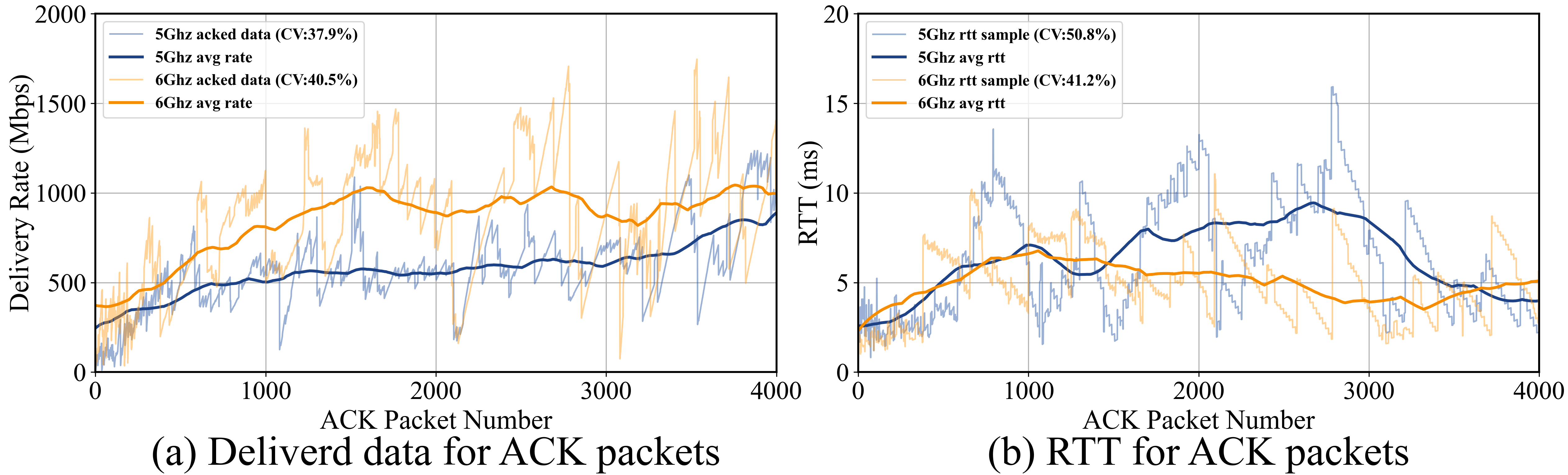}
\caption
{Statistics of ACK packets during a real MPTCP upload. The test was conducted on a dual-band, dual-link Wi-Fi network supporting 5GHz and 6GHz.}
\label{fig2}
\vspace{-0.1cm}
\end{figure}

\begin{figure}[H]
  \vspace{-0.5cm}
  \centering
  \includegraphics[width=\columnwidth]{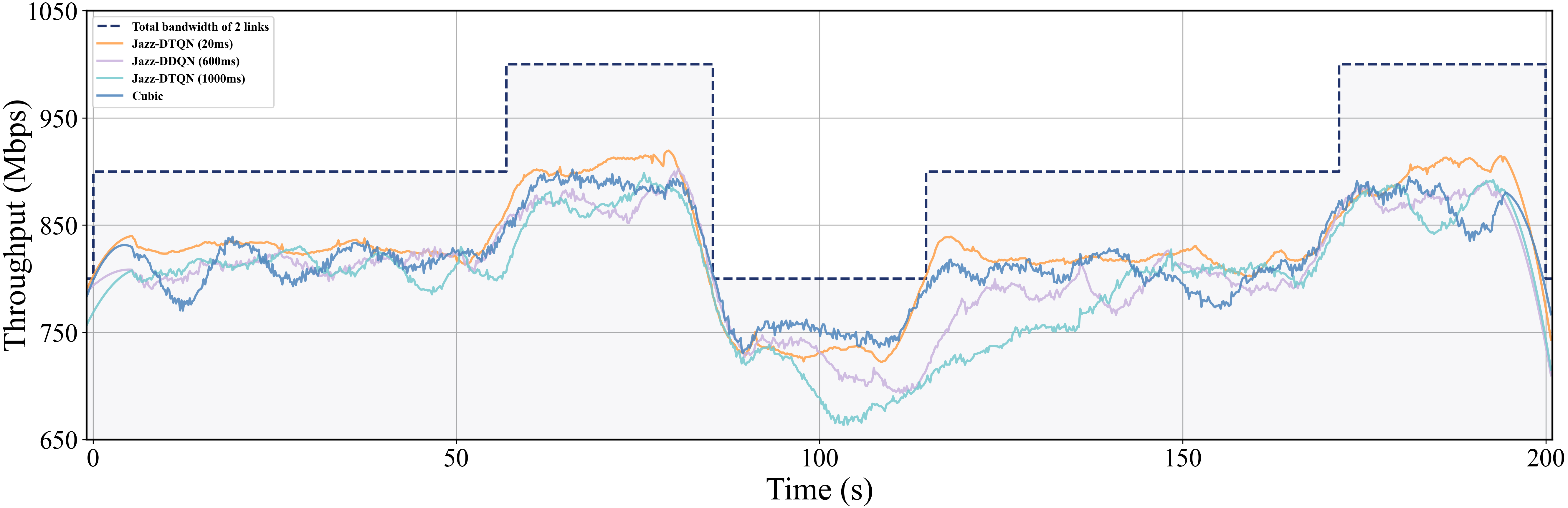}
  \caption
  {Responsiveness with different control intervals. Jazz-DTQN (1000ms) uses artificially introduced delay. The dashed line shows the total bandwidth (altered via Linux TC) of paths.}
  \label{fig3}
  \vspace{-0.2cm}
  \end{figure}
\subsection{Partial Observability Challenge and Solutions}
A challenge for DRL-based CC is forming a reliable feature representation. Relying on individual ACKs to measure instantaneous throughput and delay can be inaccurate. Phenomena like ACK aggregation (burst of delivered packets) create highly variable and misleading measurements. A simple test on a dual-band Wi-Fi multipath network showed that during an MPTCP upload using BBR (as in Fig. \ref{fig2}), both the 5 GHz and 6 GHz links exhibited a high coefficient of variation (CV) in delivered data volume and RTT. The instantaneous rate measurements can be over 40\% higher than the actual average. 

Sharp fluctuations in the data volume and RTT derived from each ACK make it difficult for the agent to form a stable assessment of the current network condition or to reliably track state transitions. This mismatch between high-frequency measurement noise and slower, true network dynamics transforms the control task into a challenging Partially Observable Markov Decision Process (POMDP) \cite{spaan2012partially}. In this context, the control loop is forced to react to sub-millisecond-level measurement noise, while the true network dynamics (e.g., changes in available bandwidth) evolve over longer timescales. An agent unable to distinguish these transient fluctuations from persistent trends based on a sequence of misleading observations will inevitably make erratic decisions, leading to suboptimal network behavior.

A straightforward approach to mitigate the problem of noisy feature representation is to have the agent observe network feedback over an extended time window. This method attempts to approximate a fully observable MDP, the model upon which single-step RL algorithms rely, where each experience is a single transition tuple: (state, action, reward, next state). In the context of CC, the agent waits for a period after each action until the network stabilizes. Aggregated metrics from this interval then form a more robust ``next state" and its corresponding reward. This strategy makes the agent's decisions less susceptible to short-term fluctuations by basing them on longer observational time slots. While deploying a pre-trained model in inference-only mode could also shorten the decision interval, this compromise would sacrifice the key advantage of RL, the ability to continuously learn from new experiences and adapt its decision-making strategy.

While extending the observation window is a straightforward fix, a more elegant approach is to empower the agent to analyze sequences of historical observations. This allows for finer-grained control without lengthening the decision interval. Such a task demands an architecture capable of capturing long-range temporal dependencies, making sequence-based models like the Transformer particularly suitable. By processing a trajectory of recent inputs, its self-attention mechanism dynamically weighs the significance of different points in the sequence. This enables the agent to discern underlying network trends from transient noise, facilitating proactive control decisions based on inferred system dynamics rather than reactive responses to misleading measurements.

We implemented tests to see how the action interval influences the performance. The evaluation was conducted in an emulated multipath network where the link capacity periodically fluctuated (bandwidth alternating between 400 and 500 Mbps, with a delay varying from 3 to 5 ms). As shown in Fig. \ref{fig3}, the Transformer-based agent (Deep Transformer Q-Network, DTQN \cite{dtqn}), with an average decision interval of approximately 20 ms (including training time per step), responded effectively to bandwidth changes, achieving an average throughput of 817 Mbps, outperforming CUBIC (814 Mbps). The Double DQN (DDQN) required a longer decision interval on the order of hundreds of milliseconds to ensure stable model updates, which diminished its responsiveness to network dynamics (804 Mbps). Furthermore, when we extended the decision interval of our method to around 1000 ms, its performance degraded significantly (775 Mbps).

\section{The Concept of Decoupled Network Function}
\label{sec:framework}
\begin{figure}[htbp]
  \centering
  \includegraphics[width=\columnwidth]{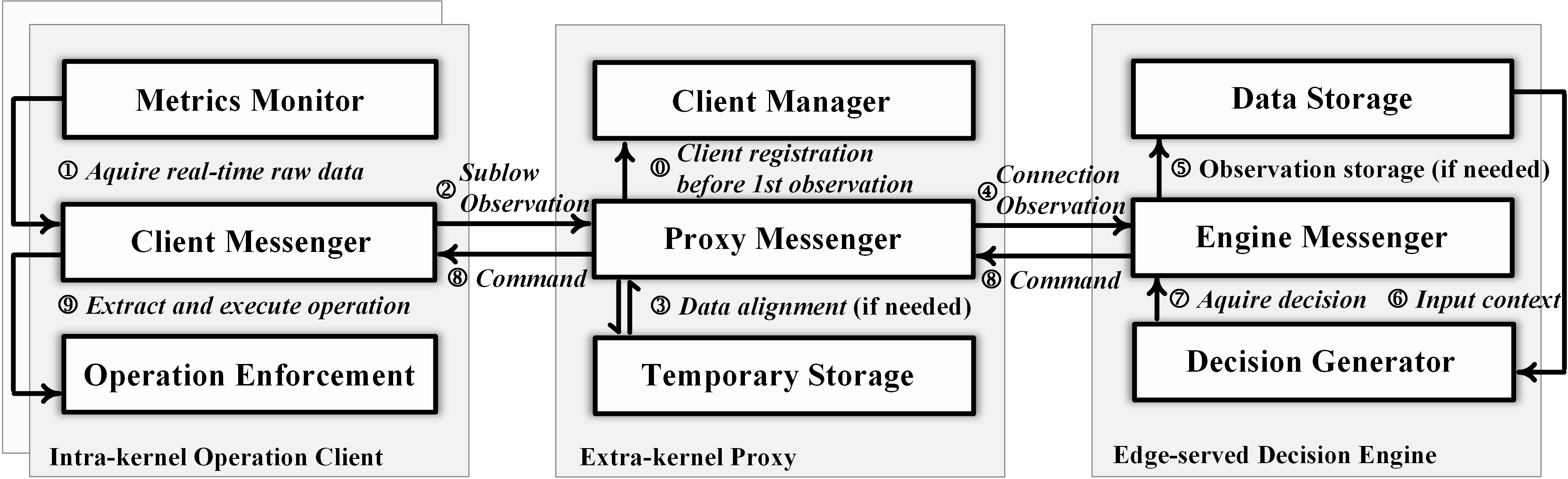} 
  \caption{The concept of network function decoupling. The decision engine may operate in the endpoint user-space. Communication methods between modules are flexible. }
  \label{fig4}
\vspace{-0.2cm}
\end{figure}

To decouple complex decision-making logic from the in-kernel datapath, we introduce a decoupled architecture, as depicted in Fig.~\ref{fig4}. This framework keeps the kernel only focused on basic metric collection and packet operations. All algorithmic complexity is offloaded to external components. This separation allows the in-kernel client to interact efficiently with external control logic via standardized interfaces, enhancing flexibility in both algorithm design and deployment. 

\subsubsection{The In-kernel Client} 
This in-kernel component is engineered to be minimal and efficient, interacting with the datapath through lightweight hooks to perform three primary tasks. First, it performs continuous monitoring of transmission status. It directly accesses kernel data structures to obtain precise metrics. The client adapts its monitoring based on specific requirements, collecting real-time data like the latest RTT for immediate decisions, or historical data such as maximum RTT over time for long-term optimization. Second, when predefined triggers (e.g., a new ACK arrival) are detected, the client formats the metrics into messages and exposes them outside the kernel via either Inter-Process Communication (IPC) mechanisms (like Netlink \cite{netlink} and Unix domain sockets \cite{ipc}) or shared memory technologies such as eBPF's shared memory maps \cite{ebpf}. The client also serves as the enforcement entity. It receives directives outside the kernel and translates these commands into direct, atomic operations. All aforementioned processes must maintain thread-safe, lock-free interactions without compromising the kernel's normal functionality. For some scenarios, kernel-user interactions may also be required to achieve sub-millisecond latency.

\subsubsection{The Ex-kernel Proxy} 
Operating in user-space, this proxy serves as the intermediary between the datapath and the decision engine (receiving data from the kernel Client Messenger and relaying commands back), enabling communication between them. It maintains connections with all client entities requesting services (e.g., maintaining a separate client for each subflow in multipath transmission). We add an invocation controller in the proxy that determines the timing of subsequent decision generation, such as triggering a new global directive upon receiving metric reports from all clients.

\subsubsection{Remote Decision Engine} 
This component receives and processes contextual data from clients through the proxy, transforming raw metrics into structured representations and generating optimization strategies. To elevate the flexibility and scalability of algorithm deployment, the decision engine can be offloaded to independent (edge) nodes, thereby leveraging software and hardware resources unavailable on the endpoint itself. By utilizing data storage, the decision engine is able to review a range of historical information, providing a more comprehensive decision-making context.


Overall, this framework provides the fundamental decoupling logic for general network functions and specific designs addressing challenges like multi-flow handling and DRL adaptation in this paper. Its principles are broadly applicable, maintaining compatibility with existing network infrastructures.

\section{Problem Formulation}
\label{sec:problem}

\subsection{Cross-Subflow Coupling}
Herein, we analyze the coupling effect based on the default \textit{minRTT} scheduler \cite{wu2021multipath}; similar coupling patterns apply to other schedulers. Consider a connection utilizing $M$ parallel subflows over paths $\mathcal{W} = \{1, 2, \ldots, M\}$. The multipath scheduler creates implicit coupling through its path selection logic. Define the scheduler's subflow availability indicator:
\begin{equation}
\label{phi}
\phi_i(t) = \mathbb{I}[cwnd_i(t) > (q_i(t) + f_i(t)) \wedge mode_i(t) \neq recovery]
\end{equation}
where $q_i(t)$ and $f_i(t)$ represent queued and in-flight packets on subflow $i$, respectively. The scheduler assigns packets to the subflow with minimum delay among available subflows:
\begin{equation}
\label{jstar}
j^*(t) = \arg\min_{i: \phi_i(t)=1} RTT_i(t)
\end{equation}

\textbf{Phenomenon (Coupling Asymmetry):} When subflow $i$ operates with excessive $cwnd_i$, queuing delay emerges as $RTT_i(t) = RTT_{min, i}(t) + Q_{queue, i}(t)$ (total RTT = propagation + queuing delay), causing the scheduler to favor alternative paths. Conversely, insufficient $cwnd_i$ triggers $\phi_i(t) = 0$, eliminating subflow $i$ from consideration. This behavior is a direct consequence of the scheduler's rule and cwnd constraint checking in (\ref{phi}-\ref{jstar}). Therefore, the traffic allocation probability (coupling effects) for subflow $i$ becomes:
\begin{equation}
\label{prob_allocation}
P(j^*(t)=i) = \frac{\mathbb{I}[RTT_i(t) = \min_{j: \phi_j(t)=1} RTT_j(t)]}{\sum_{k=1}^M \phi_k(t)}
\end{equation}

This creates asymmetric dependencies where subflow $j$'s performance is affected by all other subflows' cwnd decisions:
\begin{equation}
\label{lambda}
\lambda_j(t) = \Lambda(t) \cdot P(j^*(t)=j)
\end{equation}
where $\Lambda(t)$ is the total arrival rate and $\lambda_j(t)$ is subflow $j$'s assigned load.

\textbf{Property (Cross-flow Blindness):} Individual subflow $i$ observes state transition:
\begin{equation}
x^i_{t+1} = f_i(x^i_{t}, a^i_{t}, \lambda_i(t))
\end{equation}
where $x^i_{t+1}$ denotes the true local state for subflow $i$ (which contains complete information but is unobservable in practice), which depends on its previous true state $x^i_t$, local action $a^i_t$, and $\lambda_i(t)$. However, $\lambda_i(t)$ depends on $cwnd_j(t), \forall j \neq i$, creating hidden dependencies unobservable to local controllers. The traffic assignment $\lambda_i(t)$ in (\ref{lambda}) is a function of the global state $cwnd(t)$ and $RTT(t)$. Path $i$ can only observe local metrics $(cwnd_i(t), RTT_i(t))$ but cannot directly observe $(cwnd_j(t), RTT_j(t)), j \neq i$. Hence, from an individual path's perspective, the true system dynamics are imperceptible.

\textbf{Implication:} The Nash equilibrium of independent single-flow control is generally suboptimal compared to centralized joint control, as individual paths react to incomplete state information, necessitating joint optimization across all subflows.

\subsection{POMDP Formulation for Multipath CC}

The multipath CC problem can be formulated as a POMDP, defined as the tuple $(\mathcal{S}, \mathcal{A}, \mathcal{P}, \mathcal{R}, \mathcal{O}, \gamma)$. Here, $\mathcal{S}$ represents the state space where each $s_t = \left\{{x}^i_t \mid i \in \mathcal{W}\right\}$ contains the complete information of all subflows; $\mathcal{A}$ is the action space with $a_t = \left\{{a}^i_t \mid i \in \mathcal{W}\right\}$ representing cwnd adjustments for flows; $\mathcal{P}$ describes the state transition function (as in (\ref{prob}), state transitions of subflows are coupled via the scheduler); $\mathcal{R}: \mathcal{S} \times \mathcal{A} \rightarrow \mathbb{R}$ is the reward function designed to maximize total throughput $\sum_{i=1}^M Tput_i(t)$ subject to $\sum_{i=1}^M \max(0, RTT_i(t) - \tau_i) \leq \epsilon$, where $\tau_i$ is the delay threshold for subflow $i$ and $\epsilon$ is the violation tolerance. The agent's goal is to find a policy $\pi$ that maximizes the expected discounted cumulative reward $J(\pi) = \mathbb{E}_{\pi} \left[ \sum_{t=0}^{\infty} \gamma^t r_t \right]$, where $r_t$ is the reward at step $t$ and $\gamma \in [0, 1)$ is the discount factor; $\mathcal{O}$ represents the observation space, where each connection observation $o_t = \left\{{o}^i_t \mid i \in \mathcal{W}\right\}$ contains incomplete state information.

\begin{equation}
  \label{prob}
  \mathcal{P}(s_{t+1}|s_t, a_t) = \prod_{i=1}^M P(x^i_{t+1}|s_t, a_t, \lambda_i(t))
  \end{equation}
To address the partial observability in our problem (as discussed in Section \ref{sec:motive}), the agent processes a history of observations $(o_{t-L+1}, ..., o_t)$ to infer the underlying system state.

\begin{figure*}[t!]
  \centering
  \includegraphics[width=0.98\textwidth]{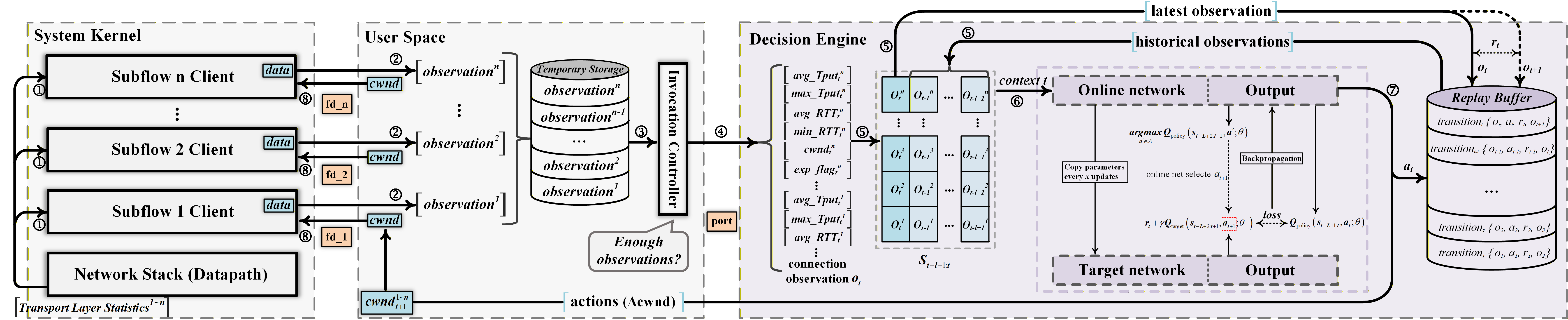}
  \caption{System Architecture for DRL-based Congestion Control. The decision engine in the architecture can accommodate DRL algorithms that optimize policies by minimizing the Bellman error, with the context sequence length set to 1 when single-step input models are employed. $\theta$ and $\theta'$ are the parameters of the online network and target network, respectively.}
  \label{fig5}
  \vspace{-0.3cm}
\end{figure*}

\section{System Overview}
\label{sec:system}
To empower multipath transmission with adaptive intelligence, we instantiate the decoupled architecture from Section \ref{sec:framework} into a fully functional system. The design (illustrated in Fig. \ref{fig5}) pivots on Bellman-error-based DRL methods that act as the ``brain". This agent processes metrics relayed from in-kernel clients and generates control directives for all subflows. This system optimizes bandwidth utilization and maintains controllable latency, adaptable to diverse networks.

\subsection{Observation Pipeline}

\begin{figure}[!b]
  \centering
  \includegraphics[width=\columnwidth]{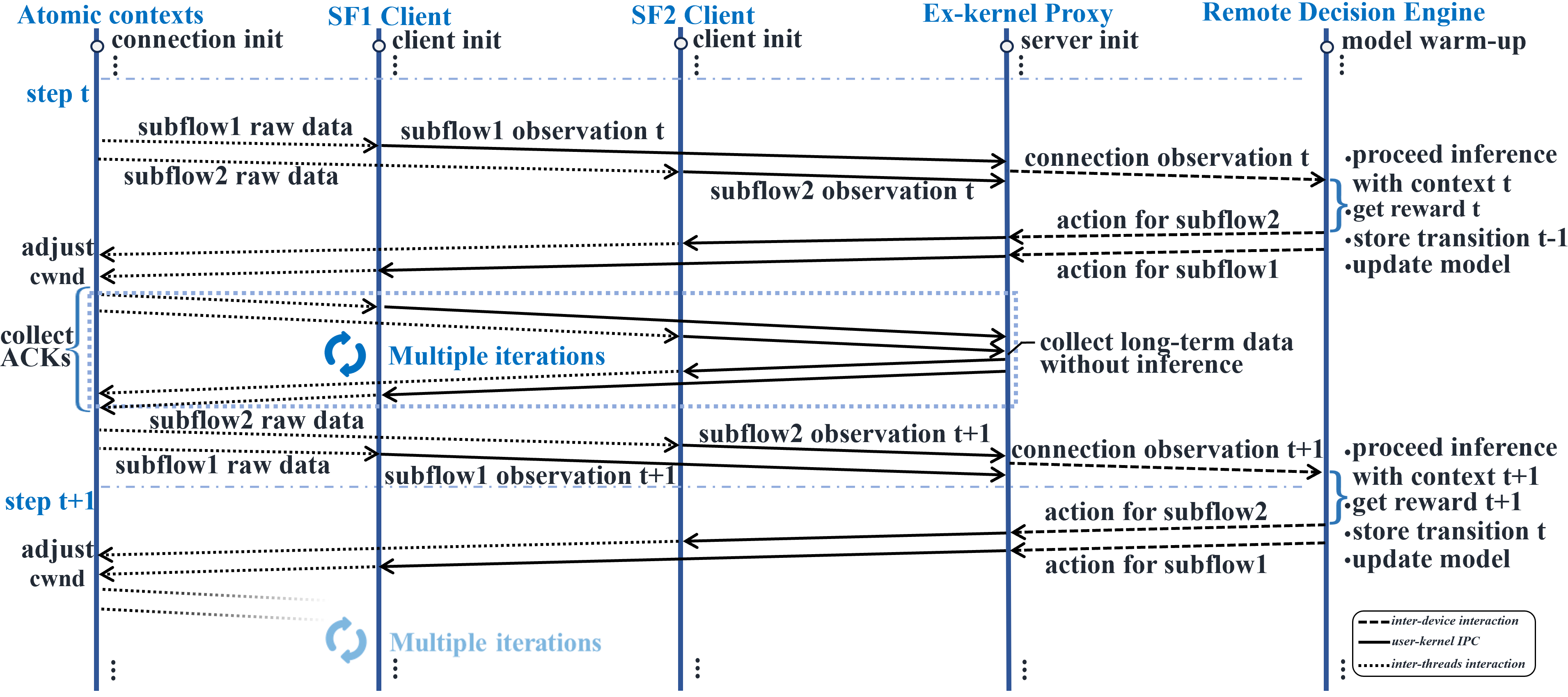}
  \caption
  {System workflow and interactions.}
  \label{fig6}
  \vspace{-0.2cm}
  \end{figure}

In our system, each subflow is managed by a dedicated client that receives directives and exposes kernel metrics, facilitating comprehensive control. These clients communicate with a decision engine through a user-space proxy for inference and updates, as depicted in Fig. \ref{fig6}.

\subsubsection{Clients' Behavior} 
Upon the successful establishment of an MPTCP connection and its subflows, each client automatically initiates operation, entering the START phase. During this initialization, in addition to reserving buffers for storing metrics and related parameters, a workqueue is scheduled to launch a dedicated background kernel thread for external interaction. This ensures that the sleepable operations remain in the process contexts instead of interrupt contexts of the kernel \cite{bharadwaj2017linux}, preserving kernel safety for atomicity and non-preemption, which are critical principles for all kernel-level operations. The START phase employs a conservative slow-start mechanism similar to traditional algorithms to achieve initial rate ramping. We intentionally simplify this phase. It automatically terminates when the cwnd reaches a low threshold and sustains stable ACK reception, letting the agent assume primary control of decisions at the earliest viable stage.

Following the initialization and the START phase, the client enters TRAIN phase and undertakes four primary functions: (i) continuous monitoring of per-ACK packet metrics such as delivered data and RTT; (ii) periodic queue draining operations to obtain propagation delay of the path; (iii) transmitting raw metrics to and receiving commands from external components through bi-directional IPC channels; (iv) enforcement of cwnd adjustments based on received directives. 

To model transmission state accurately, our client differentiates between performance and path properties. It captures per-ACK data and RTT to assess current performance, using the maximum delivery rate over a recent window as a bandwidth estimate. To measure propagation delay, the client periodically enters a PROBE phase to capture uncontended RTT. Both idle delay and bandwidth estimates are managed with finite lifecycles, prompting periodic re-evaluation of path characteristics for long-term adaptation. External I/O is handled by the background thread that continuously transmits real-time observations from the main control loop and receives target cwnd directives. The main control loop incrementally adjusts cwnd per ACK, treating the target as an upper limit.

\subsubsection{The User-space Proxy}
The proxy acts as the intermediary, bridging the kernel-level clients with the decision engine. It synchronizes the data from each subflow's client and aggregates these individual metrics into a time-aligned composite that represents a global snapshot of the entire MPTCP connection's health. To bridge the mismatched timescales between high-frequency ACK arrivals and the agent's decision-making interval, the proxy implements a short-term observation window. Within this window, it aggregates all incoming measurements to compute a smoothed representation of metrics (short-term average throughput and RTT). This mitigates the impact of transient network jitter to some extent, providing the agent with a more stable and reliable transmission representation. Specifically, the proxy also governs when to invoke the decision entity. While a single-step agent might require accumulating data across multiple windows to form a representative state, the sequence-aware model allows the proxy to trigger an inference after every single window, as the model itself leverages the historical sequence for robust decision-making. Conversely, when the agent issues a control decision, the proxy demultiplexes this command into specific cwnd directives for each subflow. They are then transmitted back to the respective kernel clients for enforcement.

\subsection{The Transformer-based Agent}
The integration of a Transformer-based agent that can decipher the non-linear interdependencies of subflow actions and adapt to time-varying channel conditions without reliance on pre-defined models. Specifically, by processing sequences of historical observations, it overcomes the partial observability inherent in single-step RL approaches. This allows the agent to learn a holistic control policy from experiential data, enabling fine-grained, context-aware policy adjustments. 

\subsubsection{Observation and Action Space}
The agent's input is a sequence of the \(L\) recent connection observations, denoted as a context \(\boldsymbol{c}_{t} = (o_{{t-L+1}}, \ldots, o_{{t}})\), where \(L\) is the sequence length, \(t\) is the time step. As shown in Fig. \ref{fig5}, each connection observation is composed of observations from all subflows (we focus on a dual-subflow scenario in this paper). The observation vector ${o}^i_t$ for a single subflow $i \in \mathcal{W}$ quantifies both flow performance and the path's underlying characteristics. We monitor its smoothed throughput, RTT, and current cwnd. The path's physical properties are determined by estimating the bottleneck bandwidth from peak delivery rates and proactively measuring the base propagation delay through periodic, controlled queue draining. Additionally, an exploration flag (expflag) is embedded in the observation (will be detailed later). The action space \(\mathcal{A}\) consists of actions for all \(M\) subflows. For each subflow, the individual action component ${a}^i_t$ of the global action $a_t$ (where $i \in \mathcal{W}$) is a discrete adjustment to its cwnd, \(\Delta \text{cwnd} = k \cdot \delta\text{, where }\delta \in \{j \mid j \in \mathbb{Z}, -n \le j \le n\}\). 


\subsubsection{Reward Shaping}
In RL, the reward signal is the primary mechanism to guide the agent towards a desired behavior. CC can be simply viewed as the problem of deciding when to increase or decrease the cwnd to maintain a reasonable queueing delay while maximizing transmission rates. Therefore, our reward function $\mathcal{R}$ is designed to teach the agent when to expand or shrink the cwnd for each subflow. We use the latency metrics as an indicator for queue depth and incorporate the expflag to incentivize the agent to probe for available bandwidth, preventing overly conservative policies.

The reward function is structured hierarchically. First, large deterministic rewards and penalties (normalized to +1 and -1, respectively) handle boundary conditions: penalties are applied when the cwnd hits its operational bounds or when actions contradict the expflag. If no boundary conditions are met, the agent receives a nuanced reward that balances throughput and latency. The expflag is set to true when the window size has not been increased for a period (6 steps in our implementation) to signal that it is time to probe for more available bandwidth. If the action aligns with the flag's indication, a positive reward is given; otherwise, a penalty is imposed.
  
In the normal regime, the reward for each subflow $i$ is a weighted sum of a throughput and a latency component:
\vspace{-0.05cm}
\begin{equation}
\label{eq:reward}
R_i = \alpha \cdot P_{D} + (1 - \alpha) \cdot R_{\rho}
\end{equation}
\vspace{-0.05cm}
where \(P_{D}\) is the RTT-related penalty and \(R_{\rho}\) is the throughput-related reward. These reward components are based on the average of historical RTT (\(\overline{D}\)) and throughput (\(\overline{\rho}\)), yielding a more stable representation of metrics. The penalty and weighting factor \(\alpha\) is calculated based on a dynamic threshold function \(T(D_{min})\) that determines when a delay increase becomes detrimental:
\begin{equation}
\label{eq:threshold}
T(D_{min}) = \frac{\beta \cdot (1 + g \cdot (D_{min} - D_{f}) / \sigma)}{D_{min}}
\end{equation}
In (\ref{eq:threshold}), the parameter $\beta$ defines the baseline permissible RTT, analogous to an acceptable queue depth. It can be tuned for specific network conditions and latency tolerances. The growth factor $g$ dictates how the threshold reacts to latency changes, while the value $D_{f}$ provides a stable RTT floor. The scaling factor $\sigma$ defines the RTT resolution for this threshold. It applies the growth factor g for each $\sigma$ µs of latency exceeding the floor $D_{f}$, letting the system treat minor RTT fluctuations within $\sigma$ $\mu$s as the same level.

As in (\ref{eq:reward}), the reward and penalty are balanced by the weighting factor \(\alpha\), which is derived by a sigmoid function:
\begin{equation}
\label{eq:alpha}
\alpha = \frac{1}{1 + e^{-\kappa \cdot (\frac{\overline{D}}{D_{min}} - T(D_{min}))}}
\end{equation}
Here, \(\kappa\) is a sensitivity parameter that controls the steepness of the sigmoid curve. The RTT penalty (as in (\ref{eq:prtt})) is proportional to how much the smoothed RTT ratio exceeds the threshold:
\begin{equation}
\label{eq:prtt}
P_{D} = -w_{D} \left( \frac{\overline{D}}{D_{min}} - T(D_{min}) \right)
\end{equation}
The throughput reward (as in (\ref{eq:rtput})) is directly proportional to the average throughput measurement:
\begin{equation}
\label{eq:rtput}
R_{\rho} = w_{\rho} \cdot \overline{\rho}
\end{equation}
where \(w_{D}\) and \(w_{\rho}\) normalize multi-scale metrics to comparable ranges. We encourage the agent to maximize stable bandwidth utilization. Finally, the total reward is the sum of rewards from all subflows, \(\mathcal{R} = \sum_{i=1}^{M} R_i\).

\subsubsection{Model Update}
The agent's policy is derived from its action-value function (Q-function), $Q(\boldsymbol{c}, a)$, which estimates the expected cumulative reward when taking action $a$ in context $\boldsymbol{c}$ and following the current policy thereafter. The agent is trained by minimizing the mean squared Bellman error from (\ref{eq:loss}). The context \(\boldsymbol{c}_{i,t}\) represents the $i$-th context sample in a training batch of size $B$, and \(a_{i,t}\) is the corresponding action. As in (\ref{eq:target}), the target value \(y_{i,t}\) is computed using the online network (parameters \(\theta\)) for action selection and the target network (parameters \(\theta'\)) for Q-value evaluation. The target calculation incorporates the immediate reward \(r_{i,t}\), a discount factor \(\gamma\), and an episode termination flag \(d_{i,t}\). An advantage of the Transformer is its ability to process entire sequences simultaneously. It computes Q-values for all \(L\) timesteps within each input trajectory in parallel, enabling efficient and holistic evaluation rather than the sequential processing of Recurrent Neural Networks (RNNs).

\begin{equation}
\mathcal{L}(\theta) = \frac{1}{B \cdot L} \sum_{i=1}^{B} \sum_{t=1}^{L} \left( Q_\theta(\boldsymbol{c}_{i,t}, a_{i,t}) - y_{i,t} \right)^2
\label{eq:loss}
\end{equation}
\begin{equation}
y_{i,t} = r_{i,t} + \gamma (1 - d_{i,t}) Q_{\theta'}(\boldsymbol{c}_{i,t+1}, \arg\max_{a'} Q_\theta(\boldsymbol{c}_{i,t+1}, a'))
\label{eq:target}
\end{equation}

\begin{figure}[htbp]
  \centering
  \includegraphics[width=0.85\columnwidth]{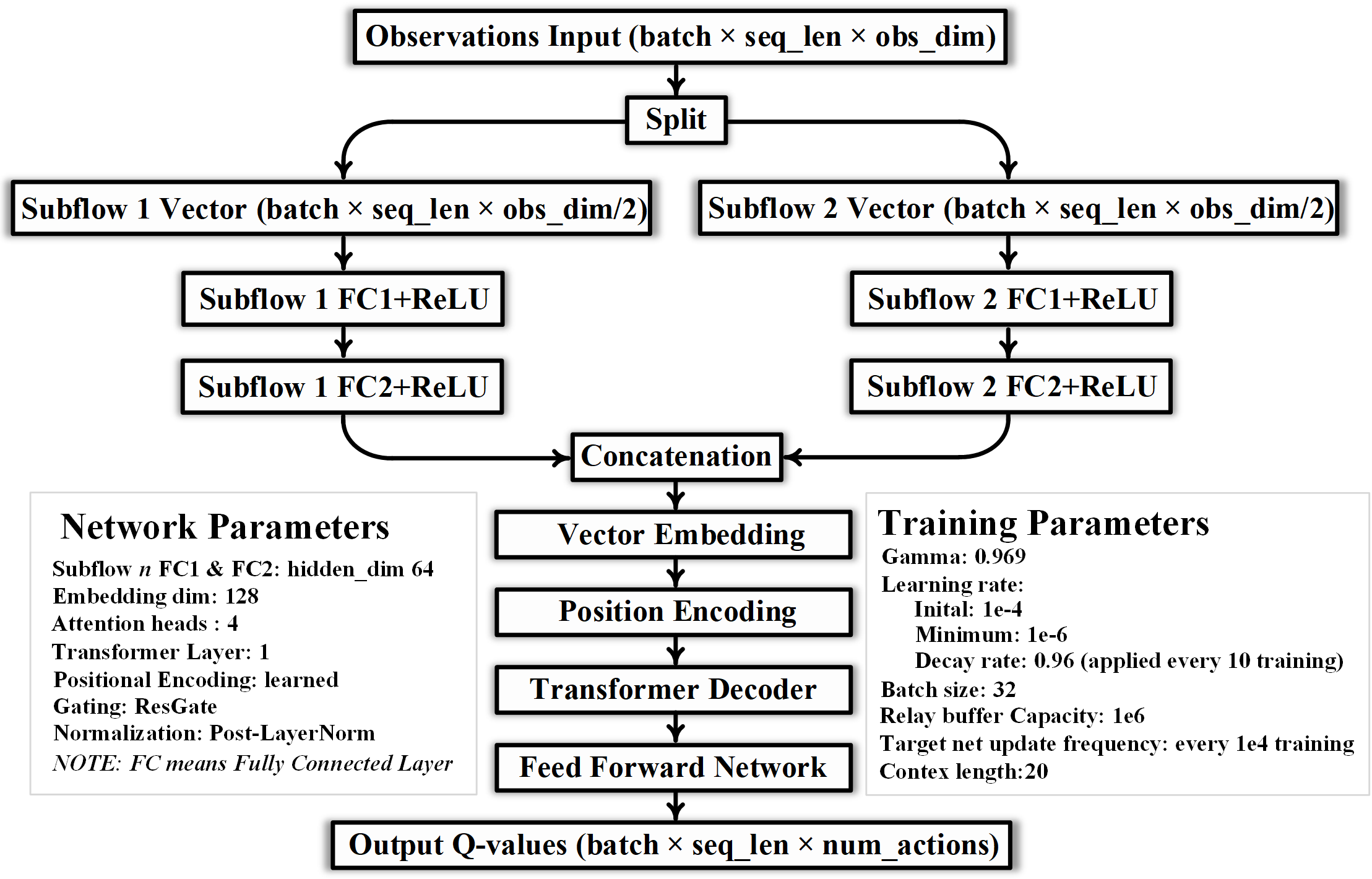}
  \caption
  {Hyperparameters and Q-network architecture.}
  \label{network structure}
  \vspace{-0.4cm}
  \end{figure}

\section{Evaluation and Validation}
\label{sec:validation}
\begin{figure}[H]
  \centering
  \includegraphics[width=\columnwidth]{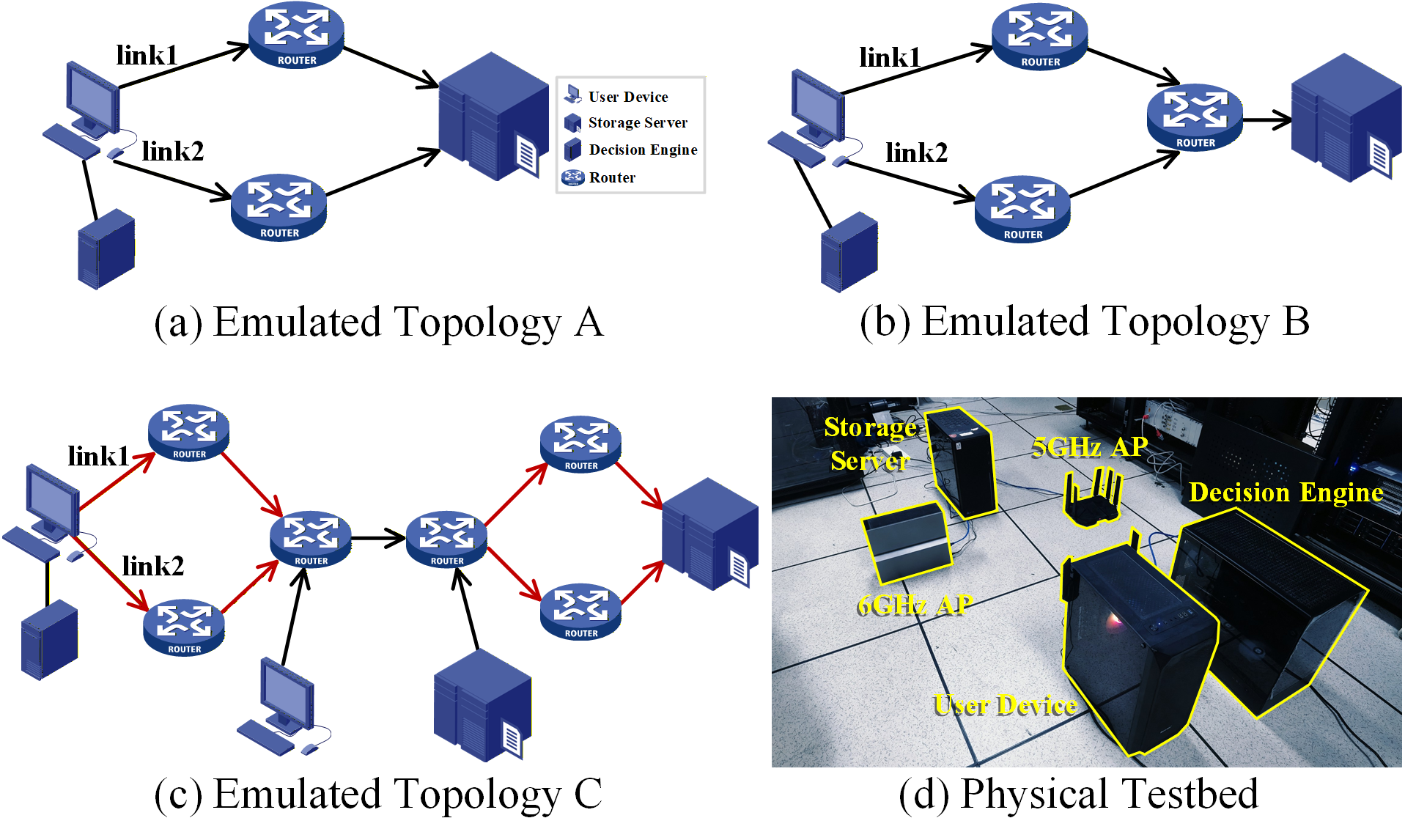}
  \caption
  {Topology of the experimental setup. The red line in (c) represents the link only traversed by multipath flows.}
  \label{fig7}
  \end{figure}

We tested our method through extensive experiments in simulated and physical environments. The hyperparameters and Q-network structure used are shown in Fig.~\ref{network structure}. The simulation setup (Fig.~\ref{fig7}(a-c)) was configured in Mininet, using a modified Linux kernel version 5.4.243. For real-world validation, our physical testbed (Fig.~\ref{fig7}(d)) employed a desktop with Mediatek MT7922 and Intel AX210 NICs, creating a dual-band (5GHz/6GHz) multipath WLAN connection to a storage server via TP-Link BE800 (6GHz) and ZTE AX5400Pro (5GHz) APs. Training and inference were accelerated using an NVIDIA RTX 4060 Ti GPU. The user device's kernel was updated to a modified Linux kernel version 6.8.0 to ensure driver compatibility for the NICs. We benchmarked Jazz against MPTCP \cite{zhao2023multipath} and TCP \cite{vielhaus2025evaluating} algorithms in both emulated and physical environments, also evaluating its performance across local and edge deployments.
\subsection{Evaluation in emulated environment}
\begin{figure}[!t]
 \centering
 \includegraphics[width=\columnwidth]{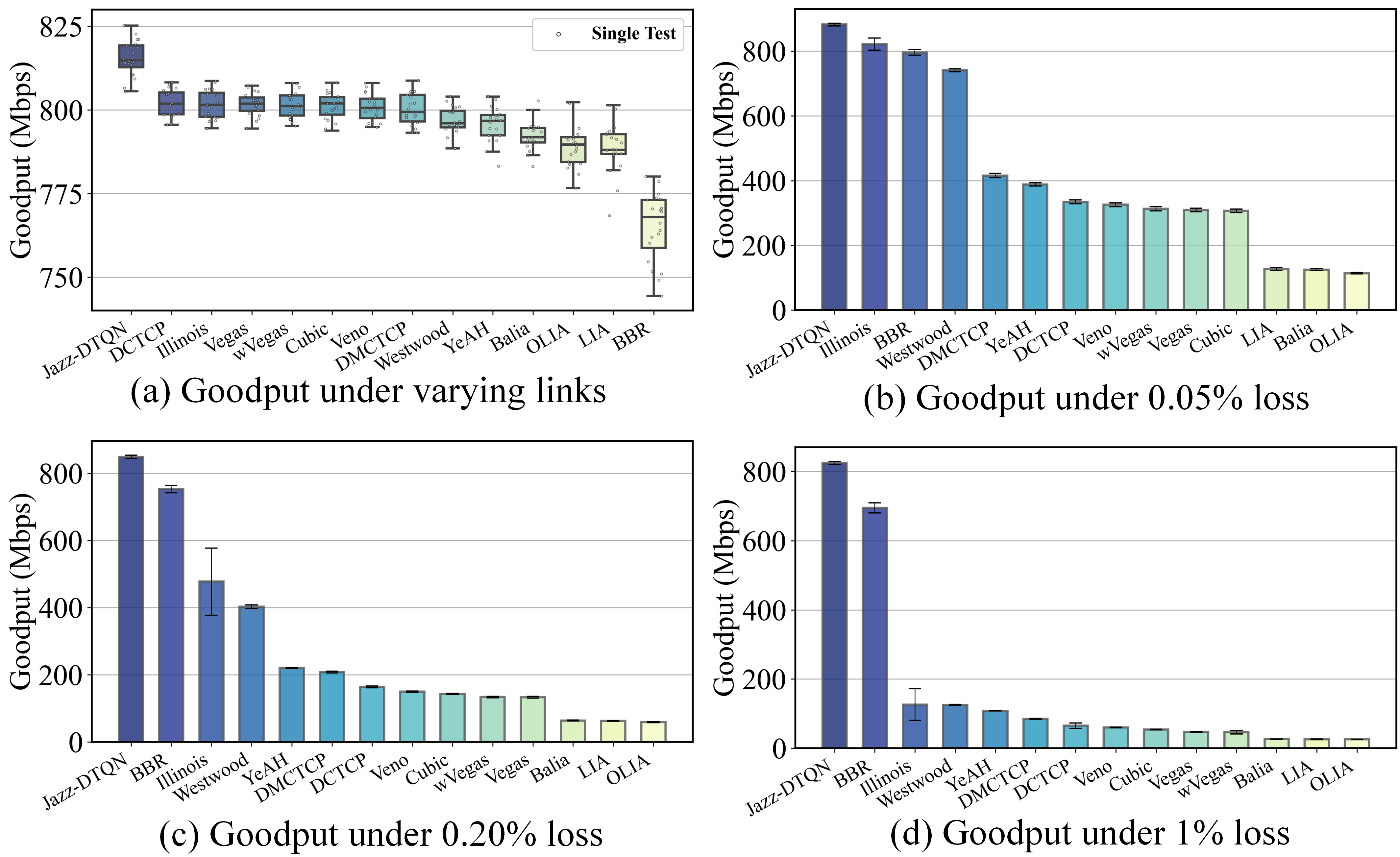}
 \caption
 {Average goodput comparison under varying links and different loss rates. The whiskers (a) and error bars (b-d) show the performance variation across multiple test runs.}
 \label{fig9}
 \vspace{-0.32cm}
 \end{figure}

We first tested bandwidth efficiency on links with fluctuating capacity (as in Fig.~\ref{fig7}(a), a setup identical to Section~\ref{sec:motive}, Fig.~\ref{fig3}), where each path's bandwidth fluctuated between 400-500 Mbps and delay varied from 3-5 ms. Each algorithm underwent 20 tests to ensure statistical robustness against link variations. The results (Fig.~\ref{fig9}(a)) showed that Jazz achieved the highest mean goodput at 815.47 Mbps. This represents a 1.75\% improvement over the next best MPTCP algorithm, wVegas (801.49 Mbps), and a 1.68\% advantage over the top-performing TCP algorithm, DCTCP (802.07 Mbps).

Within the MPTCP algorithm group, Jazz surpassed Balia by 2.92\% and the TCP-friendliness-focused OLIA by 3.34\%. In the TCP group, DCTCP, Illinois, and Vegas formed the top tier. BBR's performance was the poorest, with its mean goodput of 765.01 Mbps being 6.2\% below its group average (NOTE: we found BBR's simulated performance diverged from real-world behavior).

Using the topology in Fig.~\ref{fig7}(a) with 500 Mbps path bandwidth, Fig.~\ref{fig9}(b-d) shows performance under different packet loss rates. Jazz showed 2.8\% degradation at 0.15\% loss (882.40$\rightarrow$857.74 Mbps) and 10.7\% at 0.50\% loss (882.40$\rightarrow$787.96 Mbps). Loss-based algorithms suffered severe performance drops: Cubic declined by 44.8\% (306.14$\rightarrow$169.00 Mbps) at 0.15\% loss and 72.8\% (306.14$\rightarrow$83.23 Mbps) at 0.50\% loss. Illinois showed similar vulnerability, dropping 40.9\% and 77.3\% respectively. BBR showed moderate resilience with 4.3\% and 10.1\% reductions.


\begin{figure}[htbp]
\vspace{-0.1cm}
\centering
\includegraphics[width=\columnwidth]{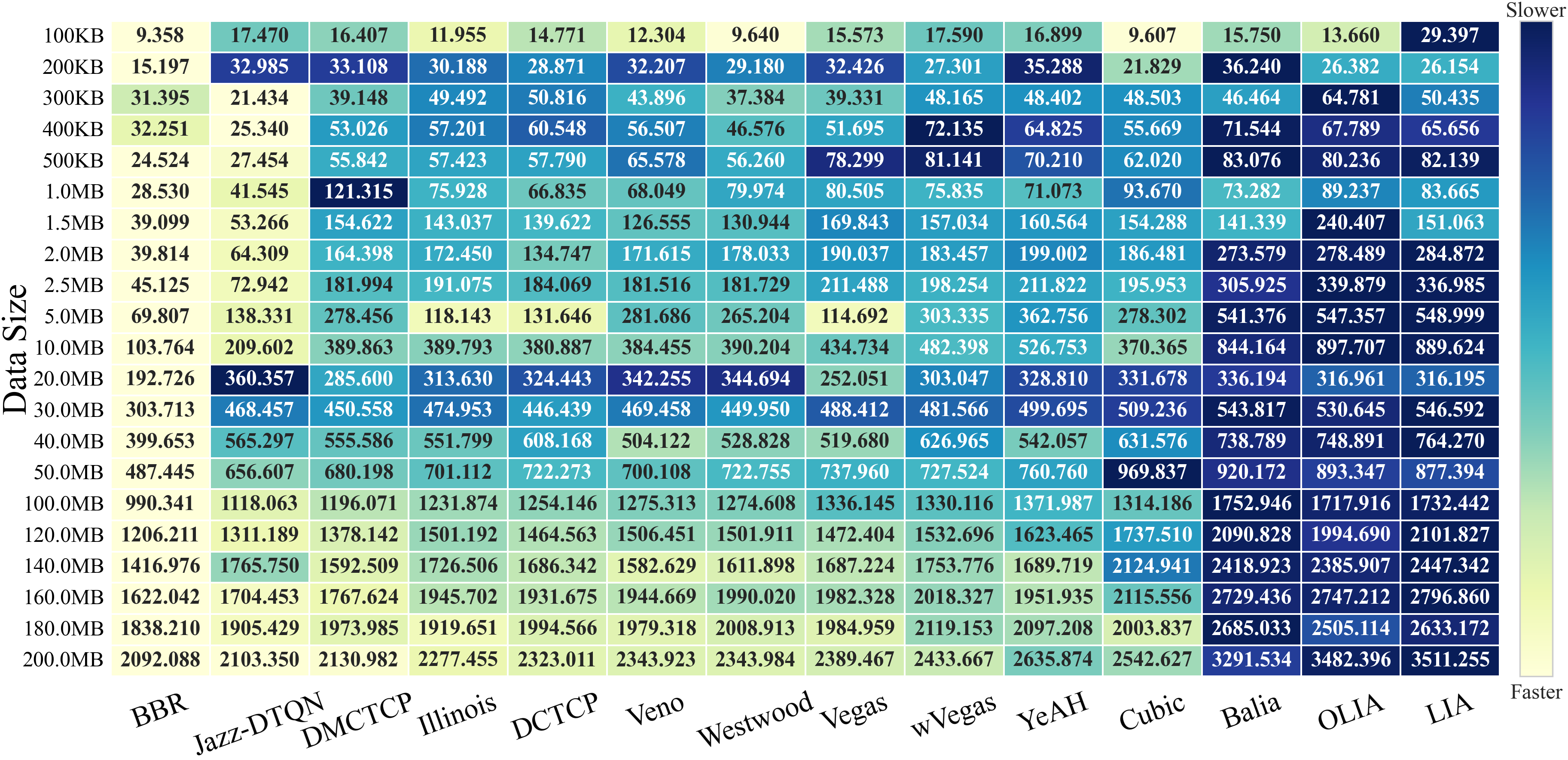}
\caption
{Average FCT (ms) comparison under varying delay.}
\label{fig10}
\vspace{-0.1cm}
\end{figure}

\begin{figure*}[!t]
  \centering
  \subfloat[{\scriptsize Goodput under varied queue size}]{\includegraphics[width=0.248\textwidth]{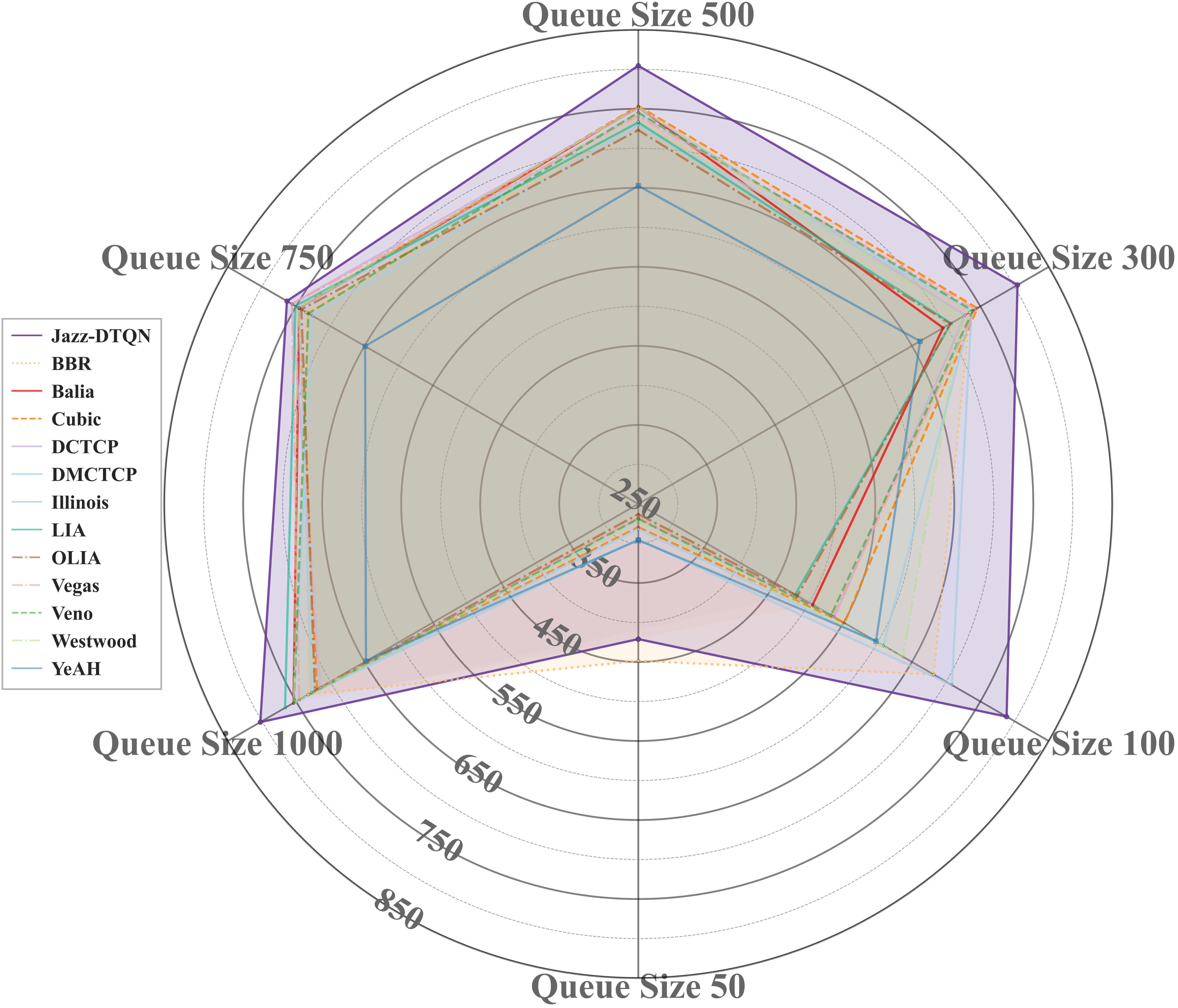}}\hfill
  \subfloat[\scriptsize FCT under 50 packets queue size]{\includegraphics[width=0.2505\textwidth]{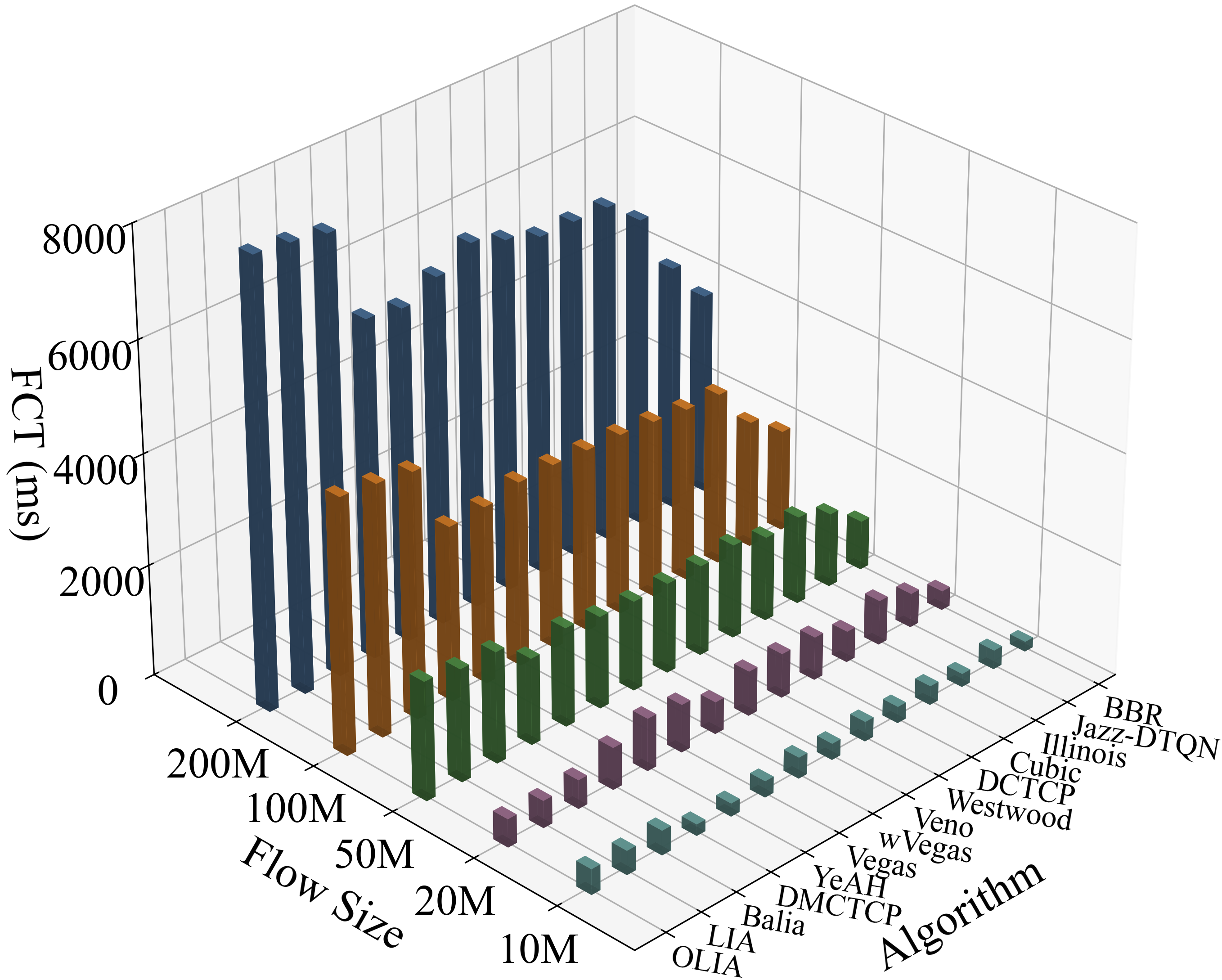}}\hfill
  \subfloat[\scriptsize FCT under 100 packets queue size]{\includegraphics[width=0.2505\textwidth]{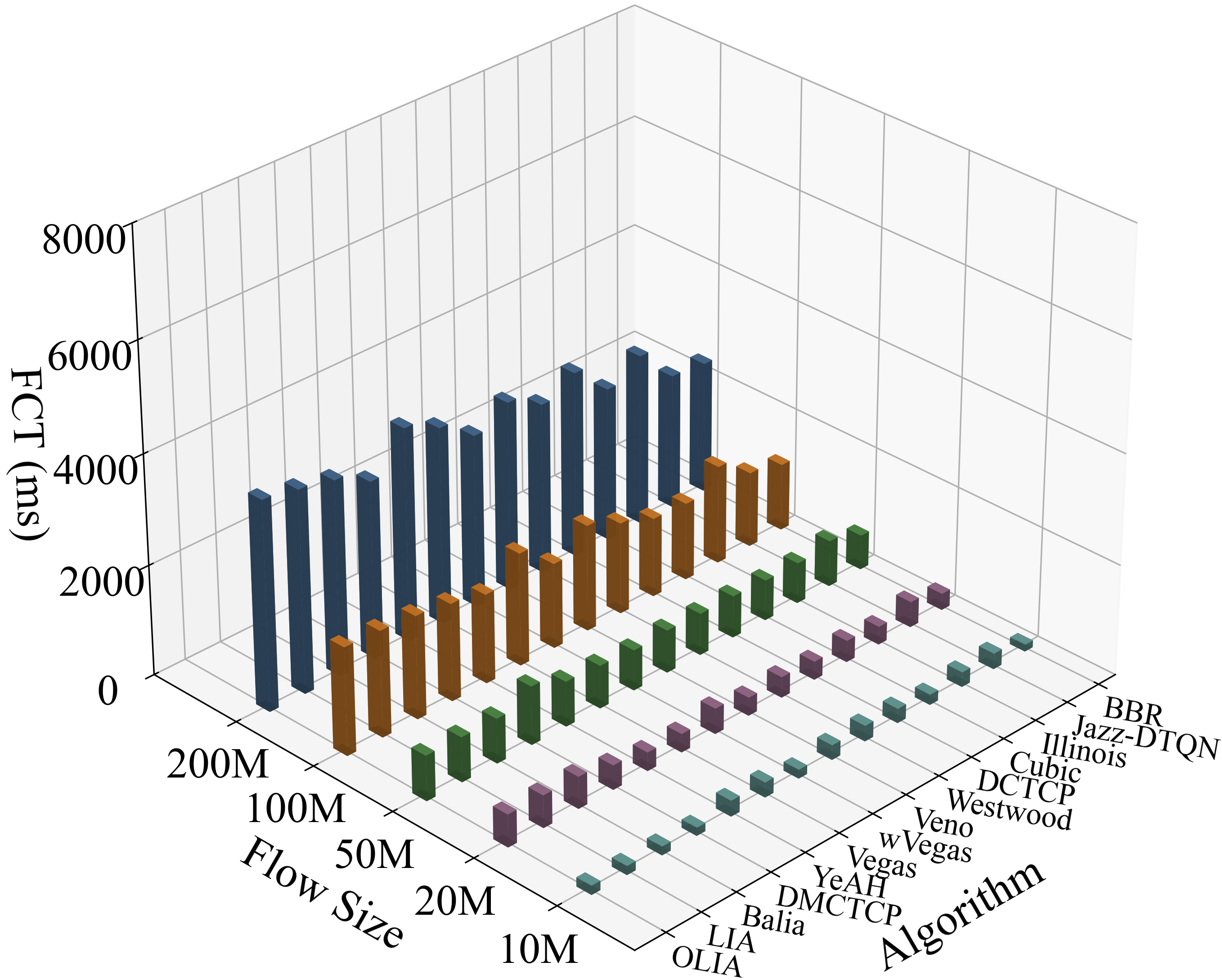}}\hfill
  \subfloat[\scriptsize FCT under 300 packets queue size]{\includegraphics[width=0.2505\textwidth]{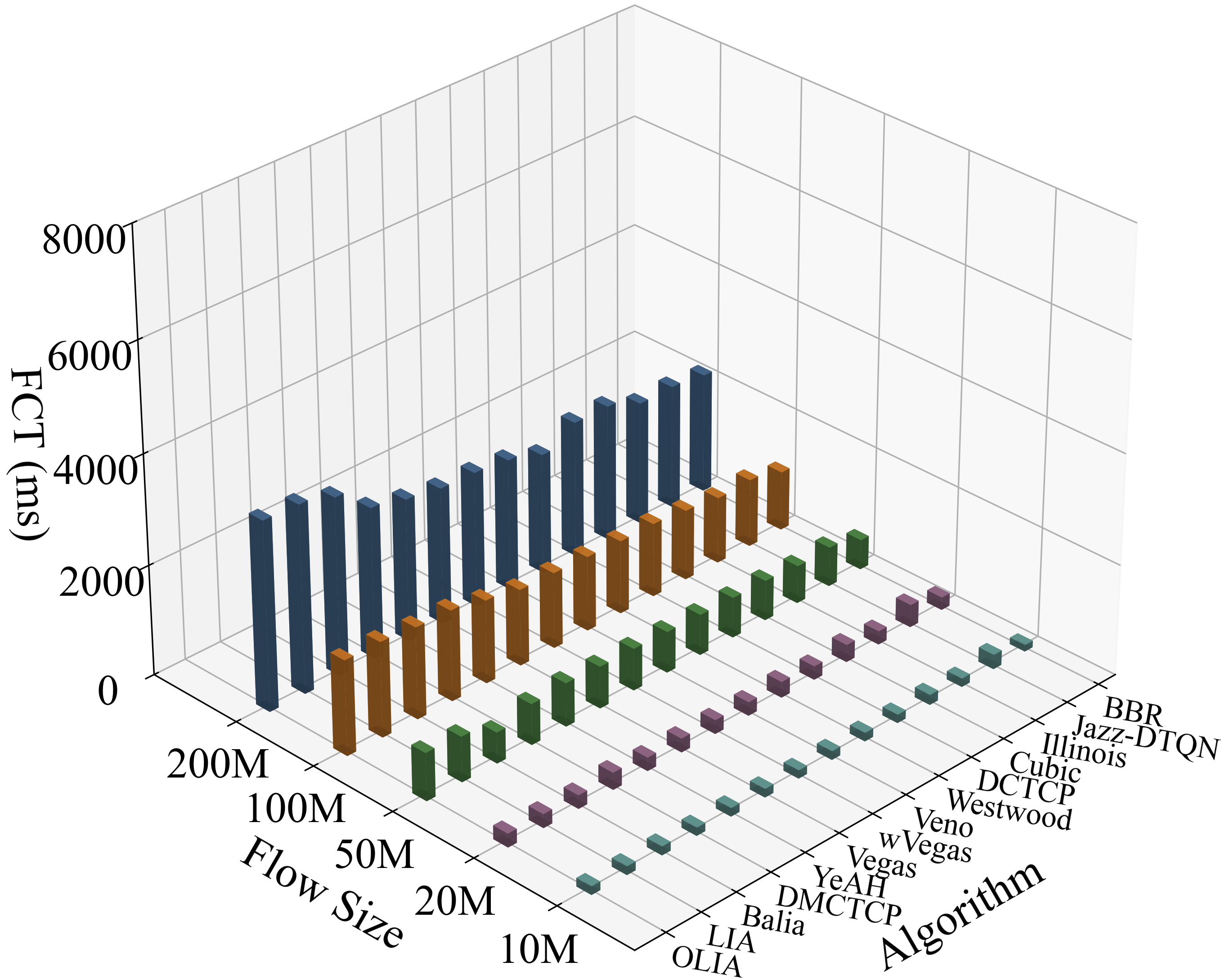}}\hfill
  \caption{Queue capacity impact on performance. The average goodput across different queue sizes is shown in (a), while (b-d) present FCT under queue sizes (we covered 10-200MB flows, given that queue constraints have little impact on shorter flows).}
  \vspace{-0.35cm}
  \label{fig11}
  \end{figure*}

To evaluate latency performance, we measured the Flow Completion Time (FCT) for short and medium-sized flows across various algorithms. The topology (Fig.~\ref{fig7}(a)) utilized a 500 Mbps bandwidth per path with 1-3 ms fluctuating delay. Each algorithm underwent 20 tests per flow size. The results in Fig. \ref{fig10} reveal a nuanced performance landscape across different flow sizes. Overall, BBR consistently achieved the lowest FCT, and DTQN-based Jazz ranked second. 

For 100KB flows, Jazz's FCT was suboptimal, which we attribute to the initial overhead of coordinating subflows that disproportionately impacts short transfers. For 1MB flows, as this overhead was amortized, Jazz became competitive, outperforming most algorithms while trailing BBR. This trend continued for larger flows (100MB), where Jazz maintained its rank as the second-fastest method, surpassed only by BBR.

We then investigated how each algorithm contended with varying buffer availability. Using the topology in Fig.~\ref{fig7}(b), we altered the bottleneck link's (1000 Mbps bandwidth) queue size to emulate buffer-to-BDP ratios from 0.3 to 3 (with buffer size of 50 to 500 packets, MTU of 1500 bytes). As in Fig.~\ref{fig11}(a), Jazz showed low sensitivity to buffer size. It reached 97\% of its peak throughput (804.52 Mbps) with a buffer of 100 packets. While BBR performed best in the smallest buffer scenario (50 packets), Jazz surpassed all algorithms at 100 packets and maintained this lead in larger buffers. Other algorithms, such as Balia, LIA, and CUBIC, heavily depend on queue capacity, showing significant throughput gains with larger buffers but poor performance in shallow queues.

Fig.~\ref{fig11}(b-d) illustrates FCT performance across varying queue capacities. Traditional algorithms like Balia, OLIA, and Westwood showed high buffer dependency, with FCT improvements of 61.9\%, 59.7\%, and 59.2\%, respectively, from 50 to 300 packet buffers. BBR and Jazz exhibited lower sensitivity, with improvements of 37.3\% and 39.6\%, indicating better adaptability to limited buffer environments. BBR excelled with small buffers (50 packets) for large flows (200MB: 3649ms vs. Jazz's 4441ms) by avoiding buffer bloat. As buffer size increased to 300 packets, Jazz reduced the performance gap for large flows (2258ms vs. BBR's 2183ms, a 3.5\% difference) while maintaining strong performance across all flow sizes.

We evaluated TCP friendliness using the topology in Fig.~\ref{fig7}(c). As shown in Table~\ref{tab:jfi}, YeAH achieved the highest fairness (JFI: 0.9998). Jazz showed moderate fairness (JFI: 0.8161), because its design focuses on bandwidth efficiency optimization rather than TCP compatibility.

\begin{table}[htbp]
\centering
\caption{Fairness of an MPTCP flow competing with an SPTCP flow, measured by Jain's Fairness Index (JFI).}
\label{tab:jfi}
\begin{tabular}{l|c|l|c|l|c}
\hline
\textbf{Algorithm} & \textbf{JFI} & \textbf{Algorithm} & \textbf{JFI} & \textbf{Algorithm} & \textbf{JFI} \\
\hline
YeAH & 0.9998 & Vegas & 0.9143 & BBR & 0.8408 \\
\hline
wVegaS & 0.9802 & BALIA & 0.9141 & Jazz-DTQN & 0.8161 \\
\hline
DCTCP & 0.9732 & OLIA & 0.9136 & DMCTCP & 0.7802 \\
\hline
Westwood & 0.9554 & Illinois & 0.8945 & CUBIC & 0.7116 \\
\hline
Veno & 0.9373 & LIA & 0.8918 & & \\
\hline
\end{tabular}
\end{table}

\vspace{-0.05cm}
\subsection{Validation in Physical Testbed}
We conducted real-world tests using the dual-band Wi-Fi testbed shown in Fig.~\ref{fig7}(d) to validate our approach beyond simulations. Besides comparing Jazz against traditional TCP algorithms (other MPTCP CCs were omitted in physical tests due to kernel/MPTCP version compatibility issues), we specifically evaluated: (1) edge entity versus local endpoint deployment performance, and (2) sequence-based versus single-step inference models (DDQN).

\begin{figure}[H]
  \centering
  \includegraphics[width=\columnwidth]{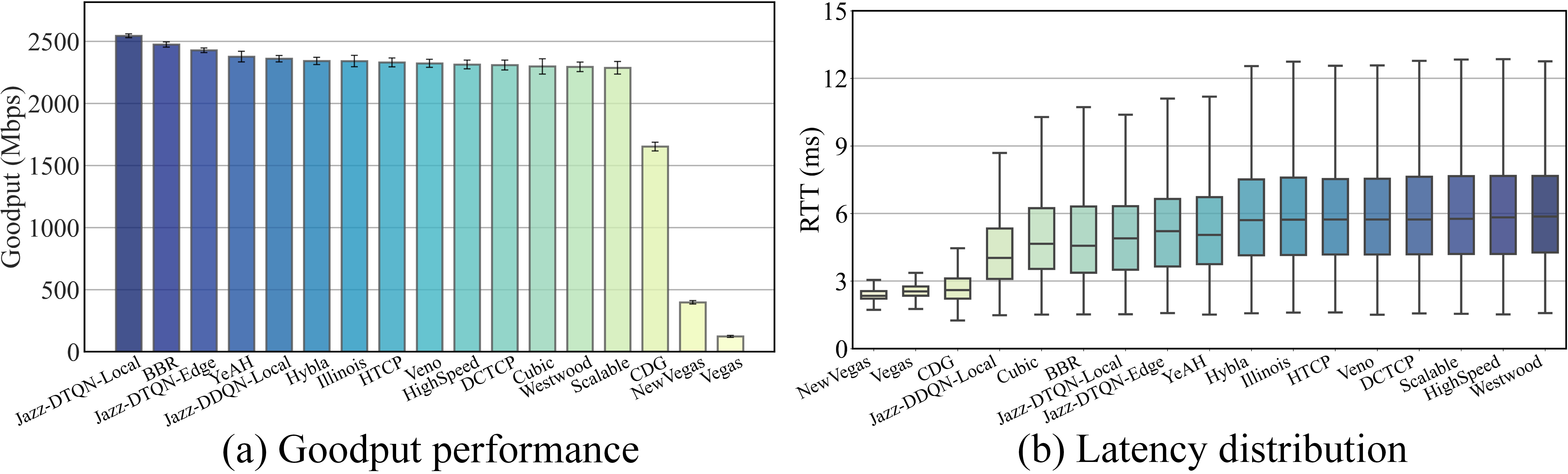}
  \caption
  {Goodput comparison in physical testbed. The error bars in (a) represent the standard deviation of the average goodput over multiple test runs, while the whiskers in (b) show the distribution of per-packet RTT.}
  \label{fig12}
  \vspace{-0.2cm}
  \end{figure}

The goodput performance comparison in the physical testbed is presented in Fig.~\ref{fig12}(a). For each algorithm, at least 20 tests (each test lasts 60 seconds) were conducted, with half performed in a daytime office and the other half in a isolated server room at night (similarly for subsequent experiments). DTQN-based Jazz demonstrated competitive performance in both local and edge deployment: local deployment achieved 2545.6 Mbps while edge deployment achieved 2427.9 Mbps. The 4.8\% performance gap between local and edge deployment confirms that millisecond-level control latency introduced by remote decision-making does impact performance in high-speed multipath scenarios. The single-step DDQN-based Jazz achieved 2360.4 Mbps.

Among traditional algorithms, BBR led with 2474.9 Mbps. Hybrid algorithms such as YeAH (2377.3 Mbps) and Hybla (2342.0 Mbps) showed decent performance by combining delay-based queue management with loss-based adaptation. Loss-based algorithms exhibited varied results: Illinois (2341.6 Mbps) performed well, while CUBIC (2298.4 Mbps) showed moderate effectiveness. Notably, the delay-gradient approach CDG (1652.7 Mbps) underperformed significantly in a wireless multipath environment.

Fig.~\ref{fig12}(b) presents the per-packet RTT distribution. Delay-based algorithms exhibited the lowest RTT, with NewVegas (2.47ms mean, 0.48ms std) and Vegas (2.59ms mean, 0.39ms std) achieving minimal latency but at the cost of throughput. DTQN-based Jazz (deployed locally) maintained a balanced RTT profile (5.19ms mean, 2.23ms std) comparable to BBR (5.15ms mean, 2.40ms std), while delivering superior throughput. Edge-deployed Jazz showed a slight latency increase (5.45ms mean, 2.34ms std) due to control loop extension. The RTT stability, reflected in the standard deviation to mean ratio, was better in DTQN-based Jazz (0.43) than in BBR (0.47).

\begin{figure}[htbp]
  \vspace{-0.2cm}
  \centering
  \includegraphics[width=\columnwidth]{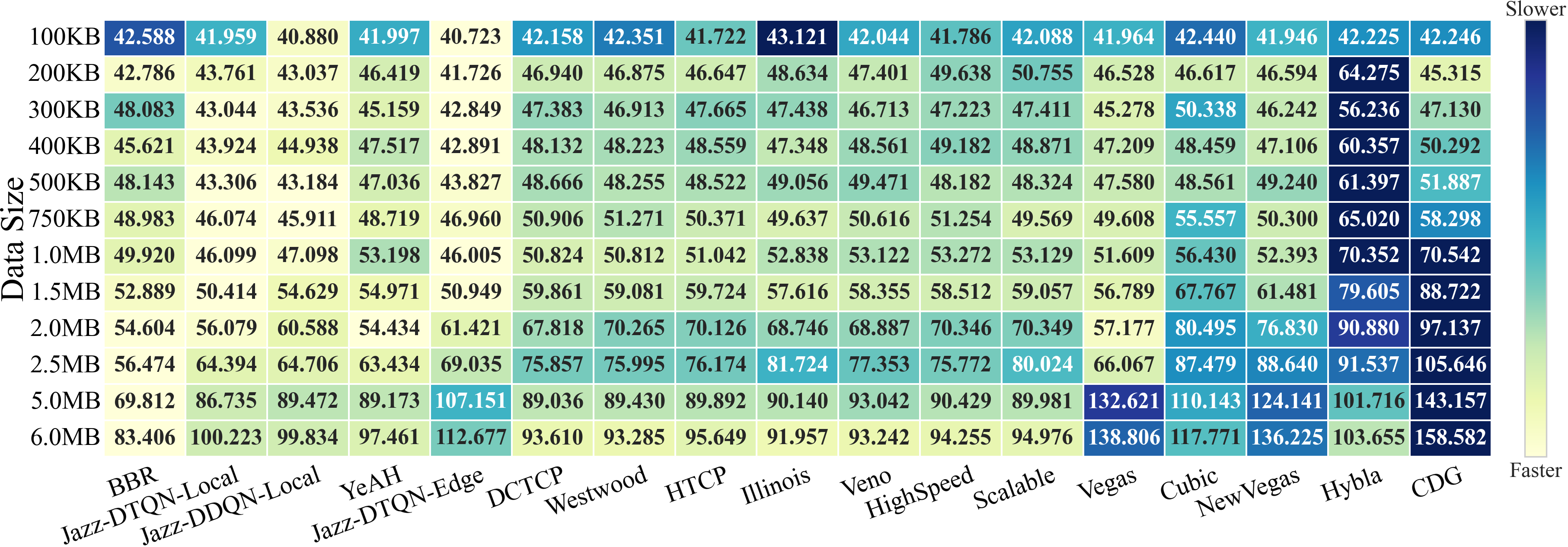}
  \caption
  {Average FCT (ms) comparison in physical testbed. }
  \vspace{-0.15cm}
  \label{fig13}
  \end{figure}

Fig. \ref{fig13} presents the average FCT comparison for short flows, where different algorithms showed varying performance across flow sizes (each algorithm underwent 20 tests per flow size). For the smallest flow (100KB), Jazz's performance was comparable to other algorithms (as for very short flows, cwnd adjustments were kernel-driven). For flows between 300KB and 1MB, our locally deployed Jazz demonstrated superior latency performance, outperforming BBR by up to 10\% at 500KB (43.31ms vs 48.14ms). For larger flows (5-6MB), BBR proved effective, achieving FCTs approximately 20\% lower than Jazz. Edge-deployed Jazz showed moderate performance due to its millisecond-level control delay.

\begin{figure}[htbp]
    \centering
    \includegraphics[width=\columnwidth]{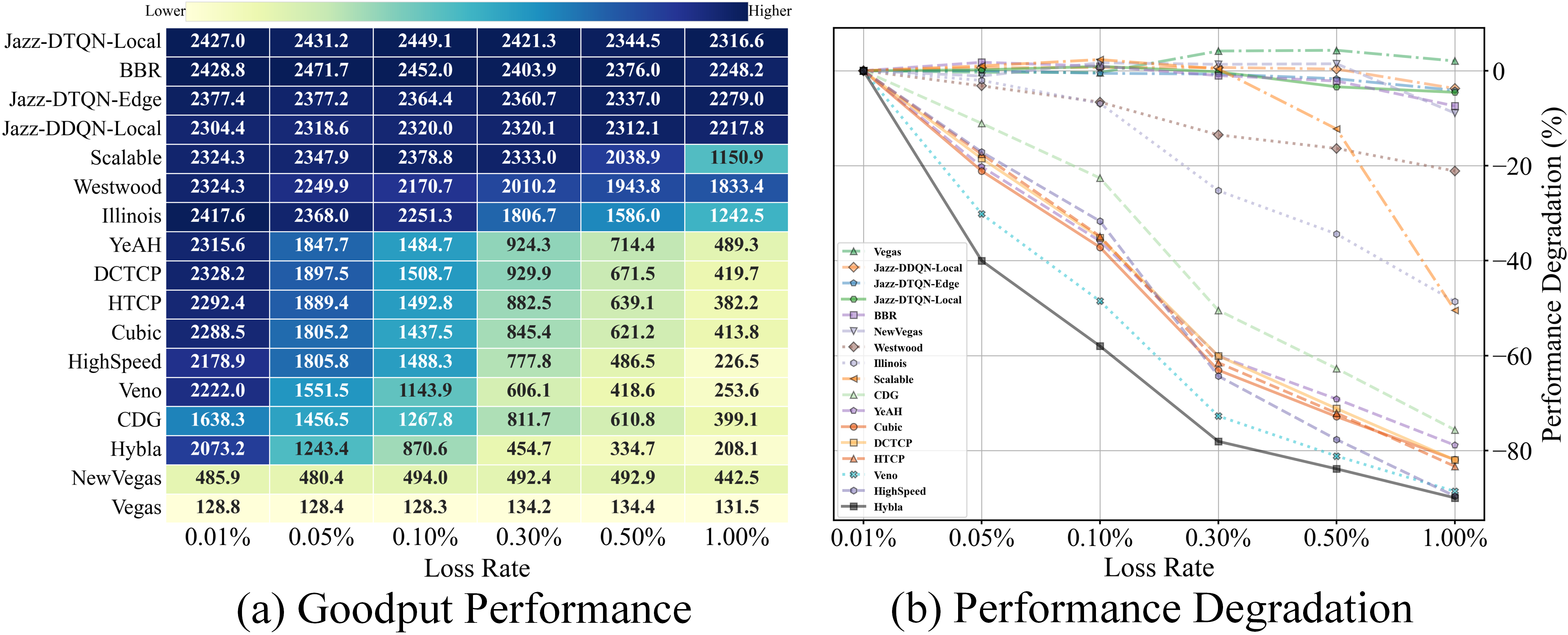}
    \caption
    {Goodput performance in varied stochastic loss. The goodput depicted in (a) is in Mbps. The packet loss rates shown represent artificial loss introduced to NICs via Linux TC, excluding the link's inherent loss.}
    \vspace{-0.5cm}
    \label{fig14}
    \end{figure}

Fig. \ref{fig14} benchmarks the throughput robustness of various algorithms against stochastic packet loss. The Jazz family showed good robustness. At a 1\% loss rate, their performance degradation ranged from just 3.8\% to 4.5\%. While loss-based algorithms like CUBIC and HTCP experienced a throughput degradation of over 80\% under the same conditions. The locally deployed Jazz sustained an average throughput of 2316.56 Mbps in 1\% stochastic loss, outperforming BBR by 3.04\% and surpassing the average of loss-based algorithms (CUBIC, HTCP, and Illinois) by 283.53\%.

\section{Related Work}
\label{sec:related}
Non-learning-based CC algorithms use varied feedback and goals. Delay-based methods \cite{copa} proactively detect congestion through RTT increases, while loss-based algorithms \cite{ha2008cubic} \cite{illinois} react to packet drops as congestion indicators. ECN-based approaches \cite{dctcp, fmptcp} leverage explicit network feedback, primarily targeting data center environments with infrastructure support. Classic multipath CC algorithms aim to address fairness and efficiency. They extend single-path principles through coupled coordination \cite{lia, olia, balia}, linking subflow controls to balance TCP-friendliness, responsiveness, and achieve load balancing across heterogeneous paths \cite{wvegas}.

For better adaptability, many works have explored learning-based CC. Early attempts employ RL \cite{qtcp} and online learning \cite{pcc} to automatically adapt transmission strategies. DRL has further advanced CC design, with works \cite{drl3r, tcp-rl} showing improved performance in various environments. For multipath scenarios, RL is applied for latency-aware path management \cite{l2mptcp}. DRL has been extended to multi-agent frameworks for subflow coordination \cite{deepcc} and adapted for multipath optimization in specialized scenarios such as distributed edge learning \cite{fair_efficient_mptcp}. These learning-based approaches, while advancing the field, share a common limitation: their reliance on instantaneous state renders them unable to capture temporal patterns in network dynamics. A promising solution involves integrating sequence-aware architectures like RNNs \cite{drqn} and Transformers \cite{dtqn} \cite{dt} into DRL, which would enable agents to learn from historical data. Furthermore, the practical aspects of deploying deep-learning-based models to interact with the in-kernel datapath are often insufficiently discussed.

Emerging technologies enable datapath extensibility and flexibility. RDMA-based solutions \cite{redn} achieve complex network offloads through self-modifying chains on commodity NICs. DPDK \cite{dpdk_elettra} provides high-performance user-space packet processing with deterministic latency for real-time applications. The eBPF frameworks \cite{emptcp, frommgen2017programming} enable safe kernel extensibility through bytecode injection. QUIC's user-space implementation represents another departure from kernel-based datapath \cite{multiflow_quic}, though performance limitations over high-speed networks merit further investigation \cite{quic_performance}. 


\section{Conclusion}
\label{sec:conclusion}

This paper demonstrates that decoupling multipath CC logic from the constrained kernel to a flexible user-space or edge node is a viable strategy for next-generation networking. We implemented this architecture with a Transformer-based DRL agent that overcomes the partial observability of single-step models by processing sequences of network observations to distinguish underlying trends from transient noise. Validated on a real-world testbed, our system achieves significantly improved bandwidth efficiency and superior robustness under loss, confirming the practicality of this approach for enabling a new class of rapidly developed, ``edge-served'' network functions. Future work will focus on optimizing cross-space communication overhead, developing generalizable congestion models, and evaluating large-scale deployment feasibility.






\begin{thebibliography}{00}
  \bibitem{rfc8684}
  A. Ford, C. Raiciu, M. Handley, et al., ``TCP Extensions for Multipath Operation with Multiple Addresses,'' RFC 8684, IETF, March 2020.

  \bibitem{narayan2018restructuring}
  A. Narayan, F. Cangialosi, D. Raghavan, et al., ``Restructuring endpoint congestion control,'' \textit{Proceedings of the 2018 Conference of the ACM Special Interest Group on Data Communication}, pp. 30-43, 2018.

  \bibitem{intel_dpdk_2014}
  Intel, ``Data plane development kit,'' 2014. [Online].

  \bibitem{Iyengar2021QUIC}
  Iyengar J, Thomson M, ``RFC 9000: QUIC: A UDP-based multiplexed and secure transport,'' \textit{Internet Engineering Task Force}, 2021.

  \bibitem{Xiao2021}
  Xiao Y, Liu J, Wu J, et al., ``Leveraging deep reinforcement learning for traffic engineering: A survey,'' \textit{IEEE Communications Surveys \& Tutorials}, vol. 23, no. 4, pp. 2064-2097, 2021.

  \bibitem{Giacomoni2024}
  Giacomoni L, Parisis G, ``Reinforcement Learning-based Congestion Control: A Systematic Evaluation of Fairness, Efficiency and Responsiveness,'' \textit{IEEE INFOCOM 2024-IEEE Conference on Computer Communications}, pp. 1451-1460, 2024.

  \bibitem{dqn}
  V. Mnih, K. Kavukcuoglu, D. Silver, et al., ``Human-level control through deep reinforcement learning,'' \textit{Nature}, vol. 518, no. 7540, pp. 529-533, 2015.

  \bibitem{attention}
  Vaswani A, Shazeer N, Parmar N, et al., ``Attention is all you need,'' \textit{Advances in Neural Information Processing Systems}, vol. 30, 2017.

  \bibitem{cardwell2016bbr}
  N. Cardwell, Y. Cheng, C. S. Gunn, et al., ``BBR: Congestion-based congestion control: Measuring bottleneck bandwidth and round-trip propagation time,'' \textit{Queue}, vol. 14, no. 5, pp. 20-53, 2016.

  \bibitem{ha2008cubic}
  S. Ha, I. Rhee, and L. Xu, ``CUBIC: a new TCP-friendly high-speed TCP variant,'' \textit{SIGOPS Operating Systems Review}, vol. 42, no. 5, pp. 64-74, July 2008. doi: 10.1145/1400097.1400105

  \bibitem{spaan2012partially}
  M. T. J. Spaan, ``Partially observable Markov decision processes,'' \textit{Reinforcement learning: State-of-the-art}, Berlin, Heidelberg: Springer Berlin Heidelberg, pp. 387--414, 2012.

  \bibitem{dtqn}
  K. Esslinger, R. Platt, C. Amato, ``Deep transformer q-networks for partially observable reinforcement learning,'' \textit{arXiv preprint arXiv:2206.01078}, 2022.

  \bibitem{netlink} 
  P. Neira-Ayuso, R. M. Gasca, and L. Lefevre, ``Communicating between the kernel and user-space in Linux using Netlink sockets,'' \textit{Software: Practice and Experience}, vol. 40, no. 9, pp. 797--810, 2010.

  \bibitem{ipc} 
  K. Wright, K. Gopalan, and H. Kang, ``Performance analysis of various mechanisms for inter-process communication,'' \textit{Operating Systems and Networks Lab, Dept. of Computer Science, Binghamton University}, 2007.

  \bibitem{ebpf} 
  M. A. M. Vieira, M. S. Castanho, R. D. G. Pacífico et al., ``Fast packet processing with ebpf and xdp: Concepts, code, challenges, and applications,'' \textit{ACM Computing Surveys}, vol. 53, no. 1, pp. 1--36, Feb. 2020.

  \bibitem{wu2021multipath}
  H. Wu, G. Caso, S. Ferlin, et al., ``Multipath scheduling for 5G networks: Evaluation and outlook,'' \textit{IEEE Communications Magazine}, vol. 59, no. 4, pp. 44-50, 2021.

  \bibitem{bharadwaj2017linux}
  Bharadwaj R, ``Mastering Linux Kernel Development: A kernel developer's reference manual,'' Packt Publishing Ltd, 2017.

  \bibitem{zhao2023multipath}
  J. Zhao, J. Liu, H. Wang, et al., ``Multipath congestion control: Measurement, analysis, and optimization from the energy perspective,'' \textit{IEEE Transactions on Network Science and Engineering}, vol. 10, no. 6, pp. 3295-3307, 2023.

  \bibitem{vielhaus2025evaluating}
  C. L. Vielhaus, C. von Lengerke, V. Latzko, et al., ``Evaluating Transport Layer Congestion Control Algorithms: A Comprehensive Survey,'' \textit{IEEE Communications Surveys \& Tutorials}, 2025.

  \bibitem{copa}
  V. Arun and H. Balakrishnan, ``Copa: Practical delay-based congestion control for the internet,'' \textit{15th USENIX Symposium on Networked Systems Design and Implementation (NSDI 18)}, pp. 329-342, 2018.

  \bibitem{illinois}
  S. Liu, T. Başar, and R. Srikant, ``TCP-Illinois: A loss and delay-based congestion control algorithm for high-speed networks,'' \textit{Proceedings of the 1st international conference on Performance evaluation methodologies and tools}, pp. 55-es, 2006.

  \bibitem{dctcp}
  M. Alizadeh, A. Greenberg, D. A. Maltz, et al., ``Data center tcp (dctcp),'' \textit{Proceedings of the ACM SIGCOMM 2010 Conference}, pp. 63-74, 2010.

  \bibitem{fmptcp}
  J. Han, K. Xue, J. Li, et al., ``FMPTCP: Achieving high bandwidth utilization and low latency in data center networks,'' \textit{IEEE Transactions on Communications}, vol. 72, no. 1, pp. 317-333, 2023.

  \bibitem{lia}
  C. Raiciu, M. Handley, and D. Wischik, ``Coupled congestion control for multipath transport protocols,'' RFC 6356, 2011.

  \bibitem{olia}
  R. Khalili, N. Gast, M. Popovic, et al., ``MPTCP is not Pareto-optimal: Performance issues and a possible solution,'' \textit{IEEE/ACM Transactions On Networking}, vol. 21, no. 5, pp. 1651-1665, 2013.

  \bibitem{balia}
  Q. Peng, A. Walid, J. Hwang, et al., ``Multipath TCP: Analysis, design, and implementation,'' \textit{IEEE/ACM Transactions on Networking}, vol. 24, no. 1, pp. 596-609, 2014.

  \bibitem{wvegas}
  Y. Cao, M. Xu, and X. Fu, ``Delay-based congestion control for multipath TCP,'' \textit{2012 20th IEEE International Conference on Network Protocols (ICNP)}, pp. 1-10, 2012.

  \bibitem{qtcp}
  W. Li, F. Zhou, K. R. Chowdhury, et al., ``QTCP: Adaptive congestion control with reinforcement learning,'' \textit{IEEE Transactions on Network Science and Engineering}, vol. 6, no. 3, pp. 445-458, 2018.

  \bibitem{pcc}
  M. Dong, T. Meng, D. Zarchy, et al., ``PCC vivace: Online-learning congestion control,'' \textit{15th USENIX Symposium on Networked Systems Design and Implementation (NSDI 18)}, pp. 343-356, 2018.

  \bibitem{drl3r}
  M. Chen, R. Li, J. Crowcroft, et al., ``RAN information-assisted TCP congestion control using deep reinforcement learning with reward redistribution,'' \textit{IEEE Transactions on Communications}, vol. 70, no. 1, pp. 215-230, 2021.

  \bibitem{tcp-rl}
  X. Nie, Y. Zhao, Z. Li, et al., ``Dynamic TCP initial windows and congestion control schemes through reinforcement learning,'' \textit{IEEE Journal on Selected Areas in Communications}, vol. 37, no. 6, pp. 1231-1247, 2019.

  \bibitem{l2mptcp}
  Y. Cao, R. Ji, L. Ji, et al., ``l2-MPTCP: A learning-driven latency-aware multipath transport scheme for industrial internet applications,'' \textit{IEEE Transactions on Industrial Informatics}, vol. 18, no. 12, pp. 8456-8466, 2022.

  \bibitem{deepcc}
  B. He, J. Wang, Q. Qi, et al., ``DeepCC: Multi-agent deep reinforcement learning congestion control for multi-path TCP based on self-attention,'' \textit{IEEE Transactions on Network and Service Management}, vol. 18, no. 4, pp. 4770-4785, 2021.

  \bibitem{fair_efficient_mptcp}
  W. Li, H. Zhang, S. Gao, et al., ``Fair and efficient distributed edge learning with hybrid multipath TCP,'' \textit{IEEE Transactions on Mobile Computing}, 2023.

  \bibitem{drqn}
  M. J. Hausknecht and P. Stone, ``Deep recurrent Q-learning for partially observable MDPs,'' \textit{AAAI Fall Symposia}, vol. 45, pp. 141, 2015.

  \bibitem{dt}
  L. Chen, K. Lu, A. Rajeswaran, et al., ``Decision transformer: Reinforcement learning via sequence modeling,'' \textit{Advances in Neural Information Processing Systems}, vol. 34, pp. 15084-15097, 2021.

  \bibitem{redn}
  W. Reda, M. Canini, D. Kostić, et al., ``RDMA is Turing complete, we just did not know it yet!'' \textit{19th USENIX Symposium on Networked Systems Design and Implementation (NSDI 22)}, pp. 71-85, 2022.

  \bibitem{dpdk_elettra}
  L. Anastasio, A. Bogani, M. Cappelli, et al., ``Integration of DPDK for Real-Time Communication in the Elettra Synchrotron Orbit Feedback Control System: Jitter and Latency Optimization,'' \textit{IEEE Transactions on Industrial Informatics}, 2025.

  \bibitem{emptcp}
  D. Shen, B. Yang, J. Zhang, et al., ``eMPTCP: A Framework to Fully Extend Multipath TCP,'' \textit{IEEE/ACM Transactions on Networking}, 2024.

  \bibitem{frommgen2017programming}
  A. Frömmgen, A. Rizk, T. Erbshäußer, et al., ``A programming model for application-defined multipath TCP scheduling,'' \textit{Proceedings of the 18th ACM/IFIP/USENIX Middleware Conference}, pp. 134-146, 2017.

  \bibitem{multiflow_quic}
  Q. De Coninck and O. Bonaventure, ``Multiflow QUIC: A generic multipath transport protocol,'' \textit{IEEE Communications Magazine}, vol. 59, no. 5, pp. 108-113, 2021.

  \bibitem{quic_performance}
  X. Zhang, S. Jin, Y. He, et al., ``QUIC is not quick enough over fast internet,'' \textit{Proceedings of the ACM Web Conference 2024}, pp. 2713-2722, 2024.

\end{thebibliography}
\end{document}